\documentclass[twocolumn,groupedaddress,aps,nofootinbib,prd,superscriptaddress]{revtex4-1} 
\usepackage{amsmath}
\usepackage{subcaption}
\usepackage{amssymb}
\usepackage{epsfig}
\usepackage{color} 
\usepackage{soul}
\usepackage[dvipsnames]{xcolor}
 \usepackage{multirow}
 \usepackage{enumitem}
 \usepackage{bbold}
 \usepackage[output-decimal-marker={.}]{siunitx}
 
\begin{document}

\title{Tetraquark systems $\bar bb \bar du$ in the static limit and lattice QCD} 

\def\UL{Faculty of Mathematics and Physics, University of Ljubljana, Ljubljana, Slovenia}
\def\IJS{Jozef Stefan Institute, Ljubljana, Slovenia} 
\author{Mitja Sadl}
 \email{mitja.sadl@fmf.uni-lj.si}
      \affiliation{     \UL } 
\author{Sasa Prelovsek}
 \email{sasa.prelovsek@ijs.si}
   \affiliation{  \UL }
    \affiliation{  \IJS } 

\begin{abstract} 
 
Two hadrons with exotic quark content $Z_b^+\simeq \bar bb \bar du$ were discovered by Belle. We present a lattice study of the $\bar bb\bar du$ systems with various quantum numbers using static bottom quarks. Only one set of quantum numbers that couples to $Z_b$ and $\Upsilon\;\pi$ was explored on the lattice before; these studies found an attractive potential between $B$ and $\bar B^*$ resulting in a bound state below the threshold. The present study considers the other three sets of quantum numbers. Eigenenergies of the $\bar bb \bar du$ system are extracted as a function of separation between $b$ and $\bar b$. The resulting eigenenergies do not show any sizable deviation from noninteracting energies of the systems $\bar bb+\bar du$ and $\bar bu+\bar db$, so no significant attraction or repulsion is found. A slight exception is a small attraction between $B$ and $\bar B^*$ at small distance for the quantum number that couples to $Z_b$ and $\eta_b\;\rho$.

\end{abstract}

\maketitle
 
\section{Introduction}\label{sec:intro}

The Belle experiment discovered two tetraquarks $Z_b(10610)$ and $Z_b(10650)$ with $J^{P}\!\!=\!\!1^+$  and $I\!\!=\!\!1$ in 2011 \cite{Belle:2011aa,Garmash:2014dhx}. Both resonances were first observed in decays to $Z_b^{\pm}\to \Upsilon (nS) \pi^{\pm}$ and $Z_b^{\pm}\to h_b (mP) \pi^{\pm}$, which indicates the exotic flavor content $Z_b^+\!\sim \! \bar{b}b\bar{d}u$. The neutral isospin partner $Z^0_b(10610)\!\sim\! \bar{b}b\bar{q}q\!\sim\!  \bar{b}b(\bar uu\!-\!\bar dd)$ was also discovered \cite{Belle:2013urd}. Later, Belle established that $Z_b(10610)$ and $Z_b(10650)$ predominantly decay to $B\bar{B}^*$ and $B^*\bar{B}^*$, respectively \cite{ParticleDataGroup:2020ssz}. Their masses are slightly above these two thresholds. Many phenomenological   studies   have been performed, for example, \cite{Wang:2018jlv,Kang:2016ezb,Ortega:2019uuk,Wang:2018pwi,Yang:2017rmm,Voloshin:2017gnc,Goerke:2017svb,Dias:2014pva,Guo:2016bjq,Ali:2011ug,He:2014nya,Karliner:2013dqa,Cleven:2013sq,Esposito:2016itg}, and the majority indicate that the $B^{(*)}\bar B^*$ molecular Fock component is essential for $Z_b$.

No lattice studies of the $\bar{b}b\bar{q}q$ resonances via the rigorous L{\"u}scher formalism are available. This is too challenging at present since one would have to determine a scattering matrix of at least seven coupled channels from a very dense spectrum of eigenenergies.

\begin{figure*}[ht!]
\begin{tabular}{cccc} 
         \multirow{4}{*}{
	\begin{subfigure}{0.28\textwidth}
	\vspace{2.0cm}
	\includegraphics[width=\linewidth]{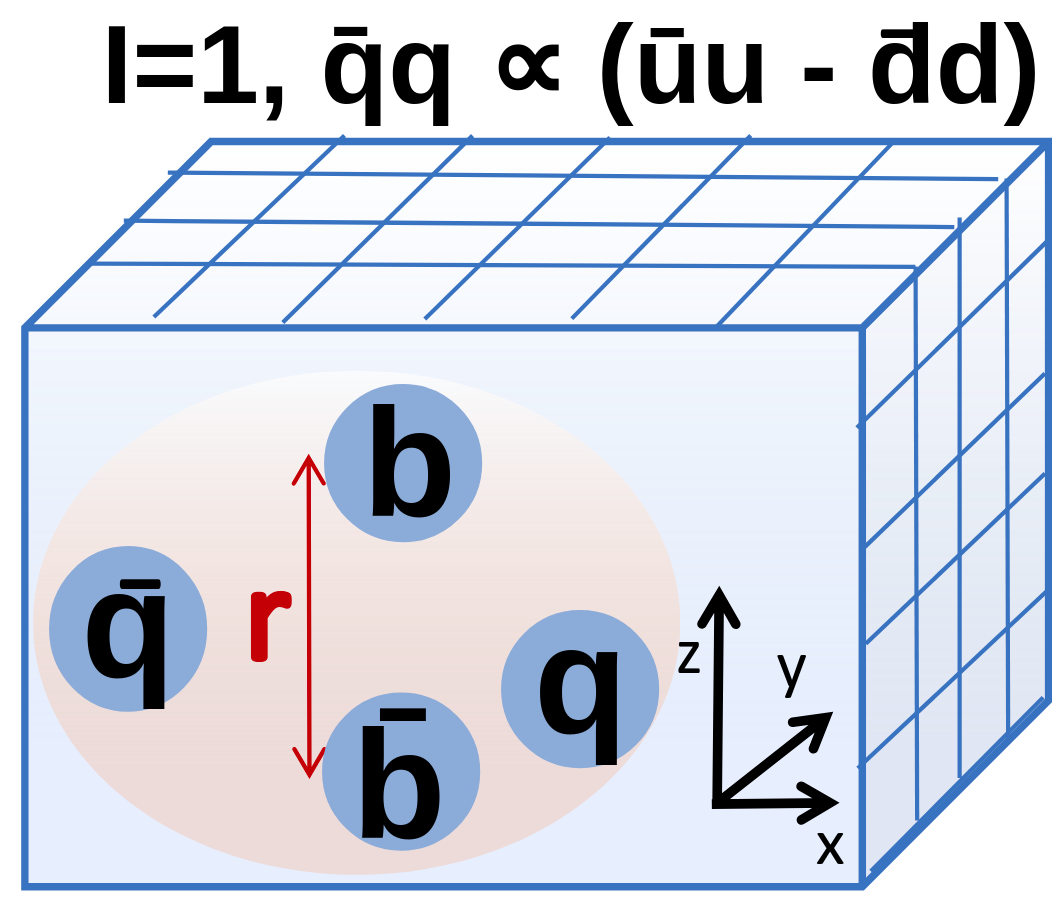}
	\vspace{2.1cm}
	\caption{ } \label{fig:1a}
	\end{subfigure}
								}         &   
	\begin{subfigure}{0.18\textwidth}
	\includegraphics[width=\linewidth]{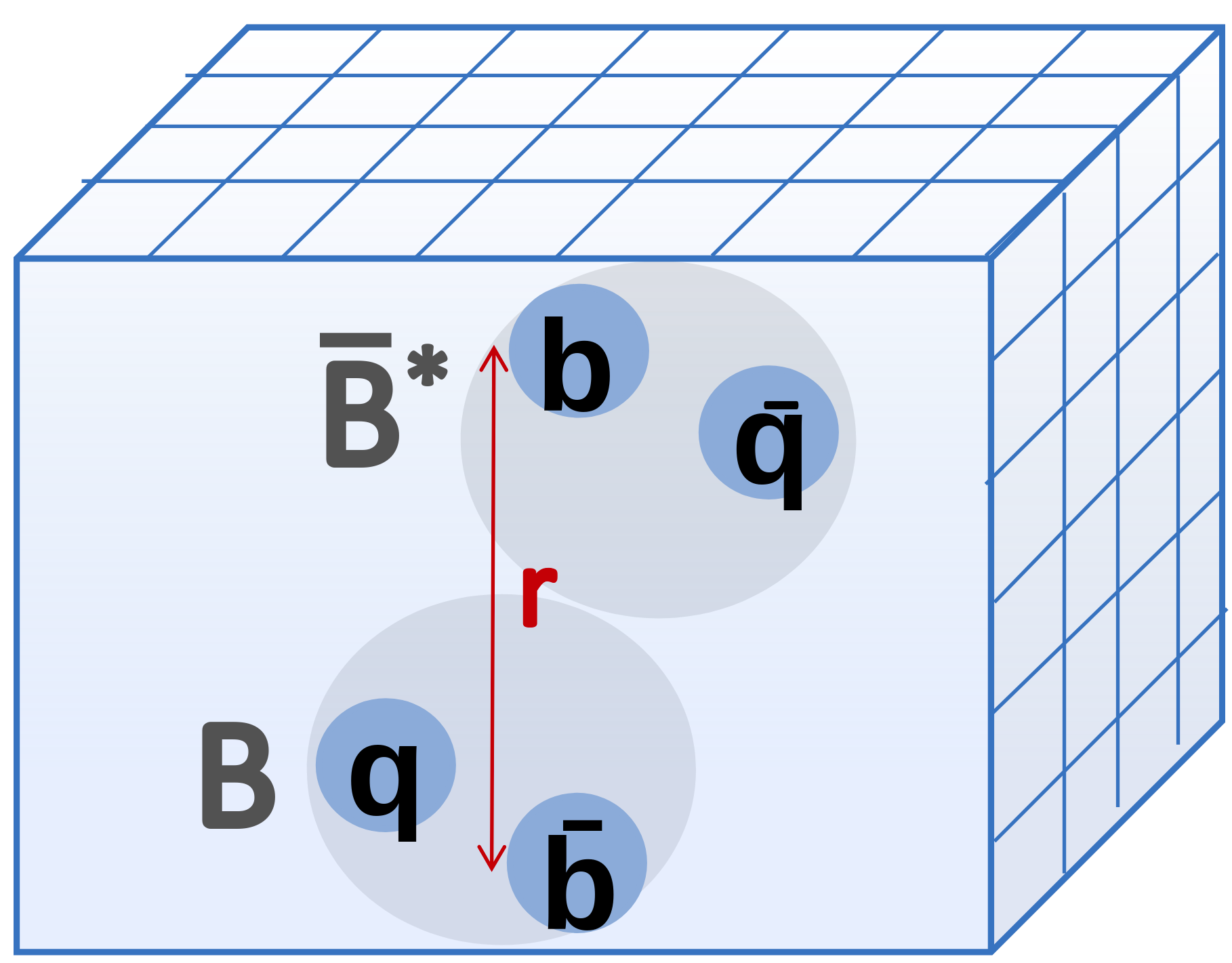}
	\end{subfigure}     
						  &   
	\begin{subfigure}{0.18\textwidth}
	\includegraphics[width=\linewidth]{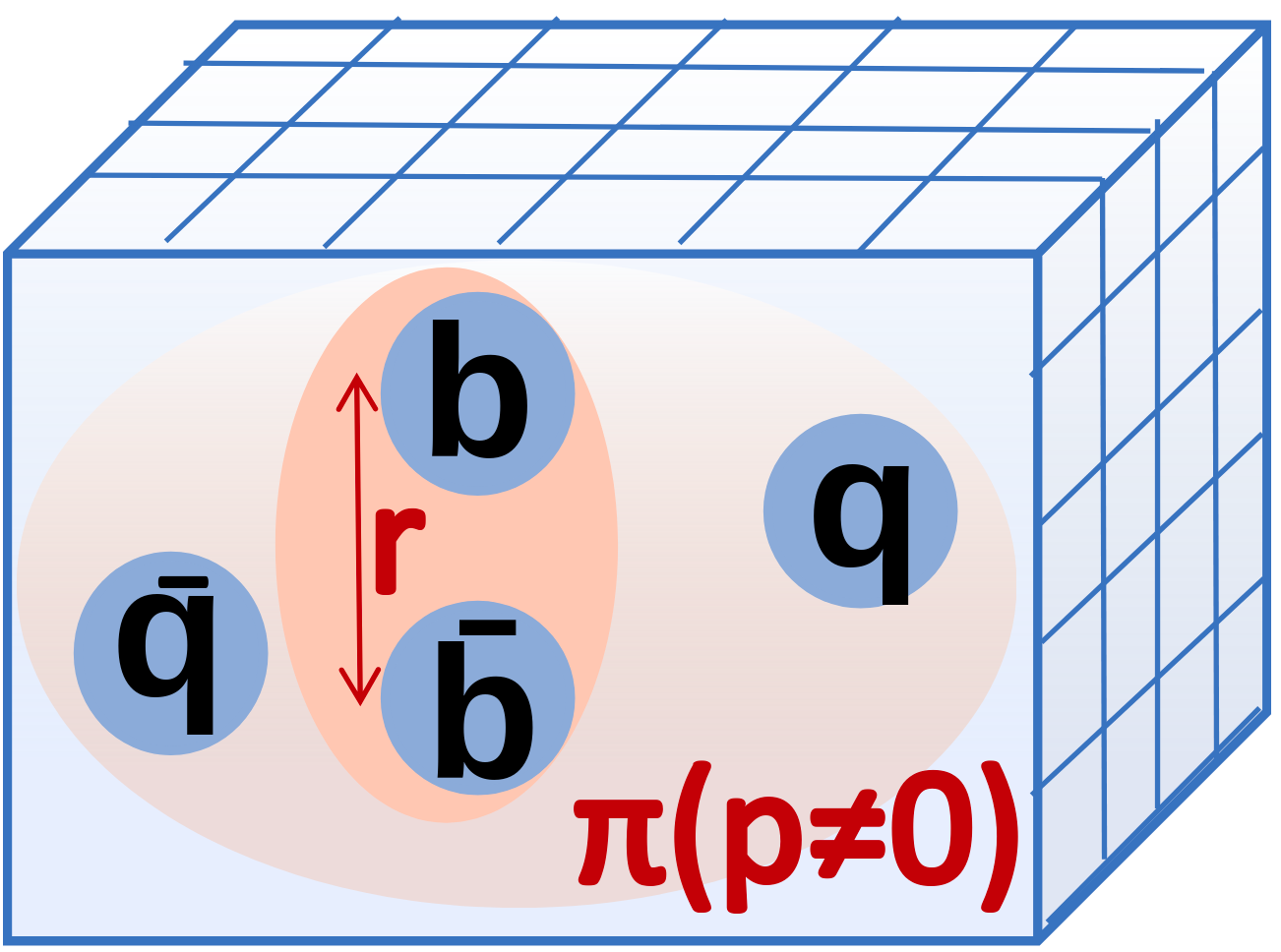}
	\end{subfigure}     
							   &   		   \\
                                     &   
	\begin{subfigure}{0.18\textwidth}
	\includegraphics[width=\linewidth]{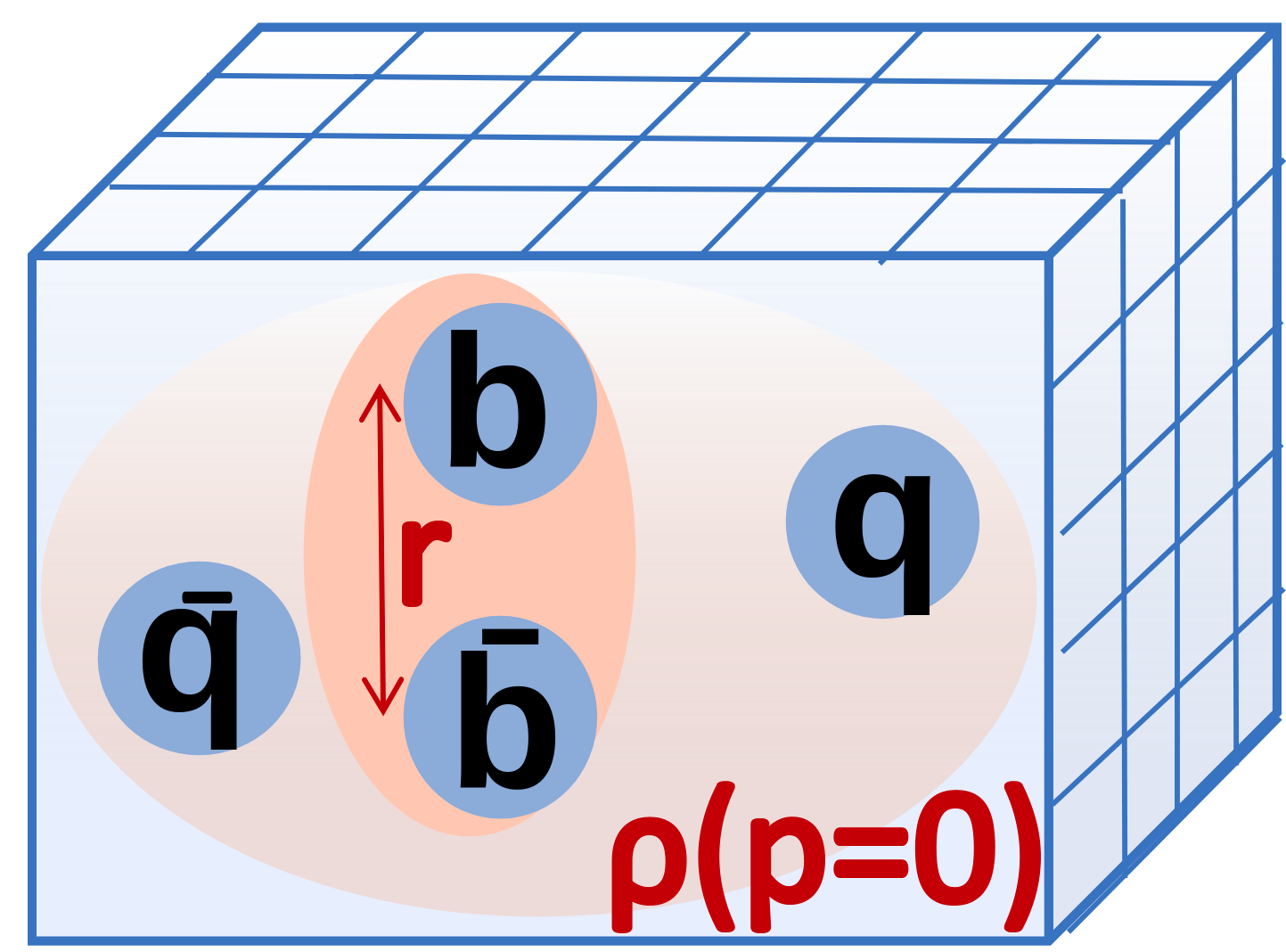}
	\end{subfigure}     
						  &   
	\begin{subfigure}{0.18\textwidth}
	\includegraphics[width=\linewidth]{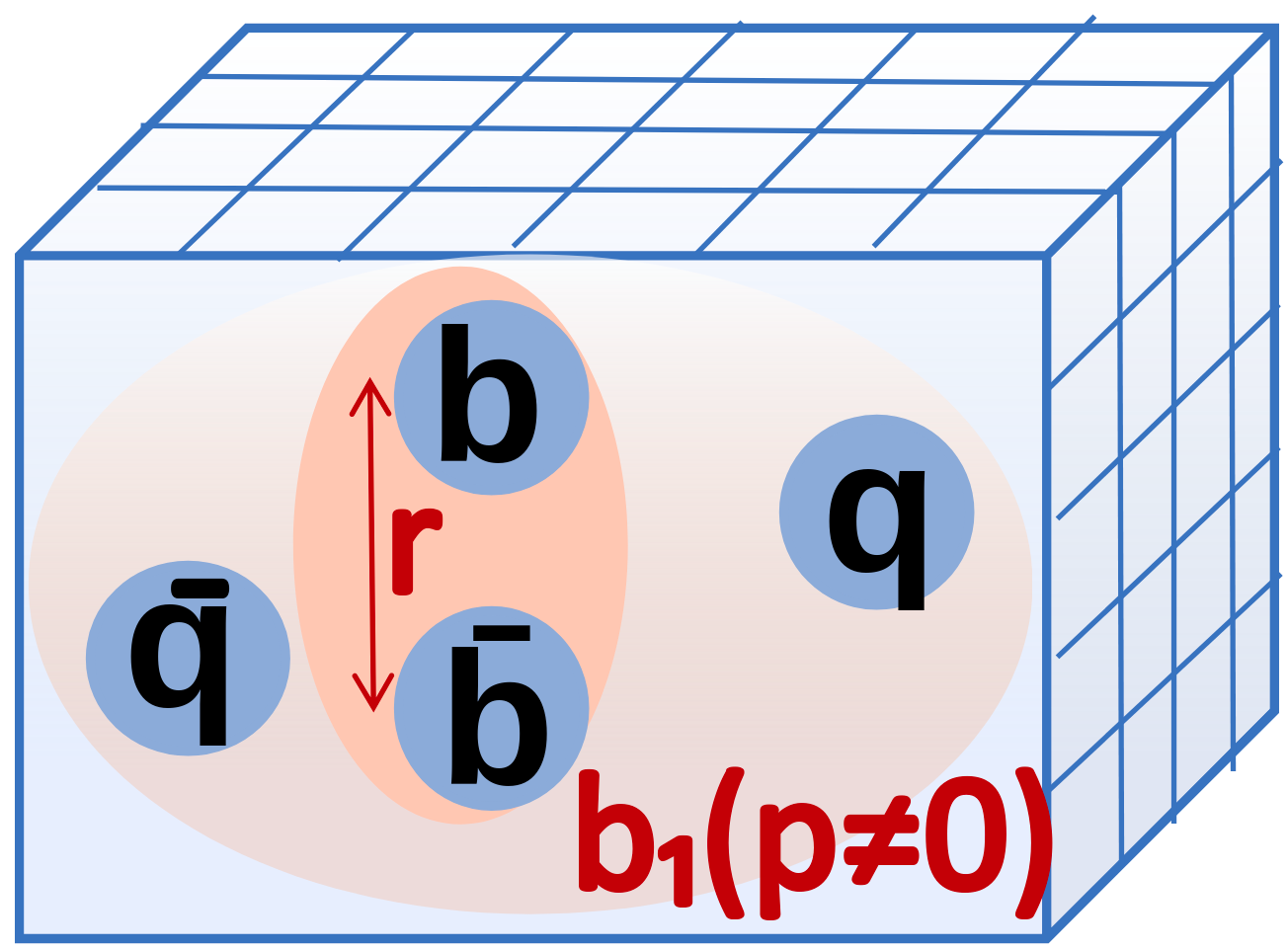}
	\end{subfigure}     
							   &    
	\begin{subfigure}{0.18\textwidth}
	\includegraphics[width=\linewidth]{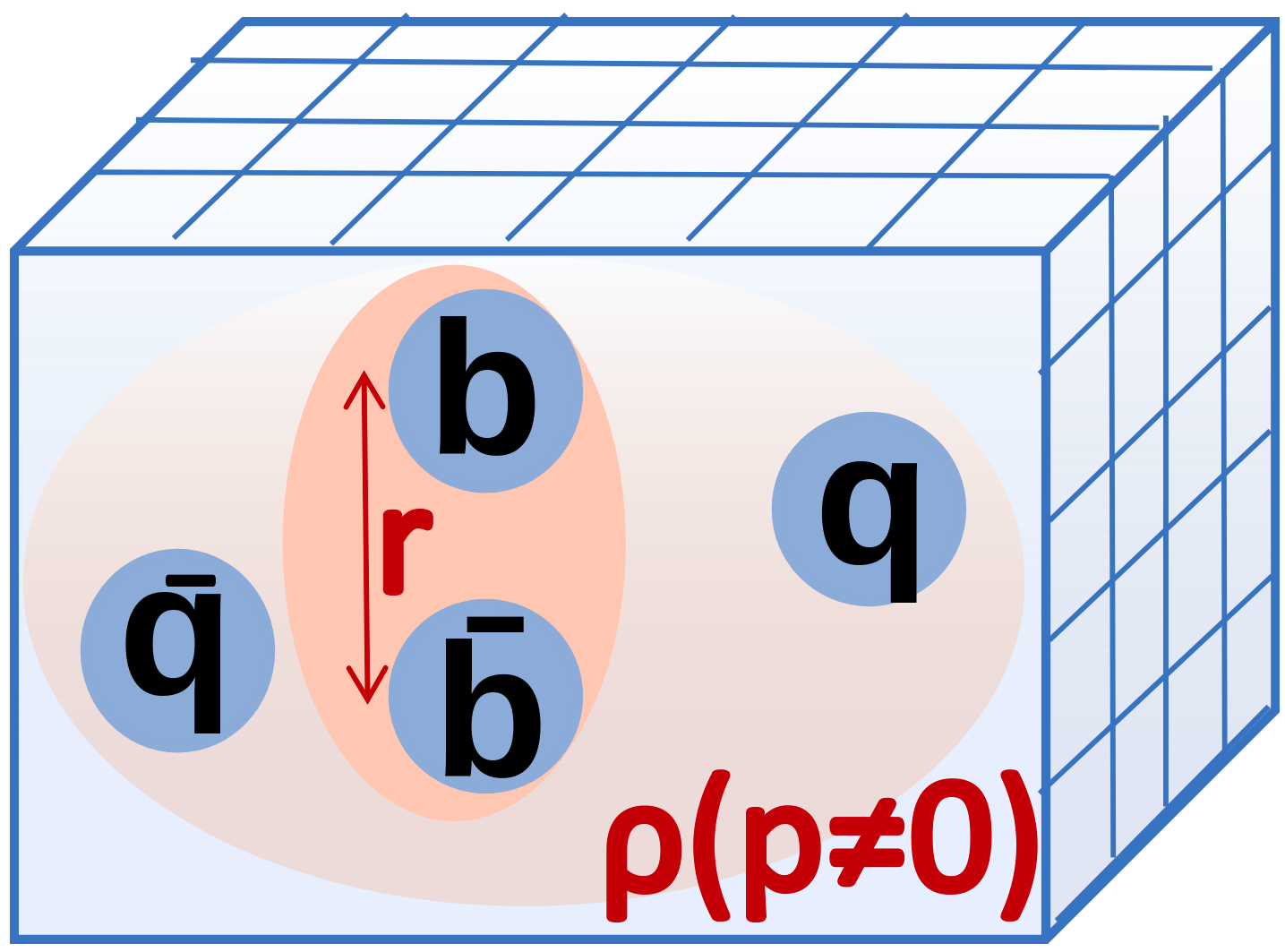}
	\end{subfigure}     
									   \\
                                     &   
	\begin{subfigure}{0.18\textwidth}
	\includegraphics[width=\linewidth]{4q_illustration_rhop.png}
	\end{subfigure}     
						  &   
	\begin{subfigure}{0.18\textwidth}
	\includegraphics[width=\linewidth]{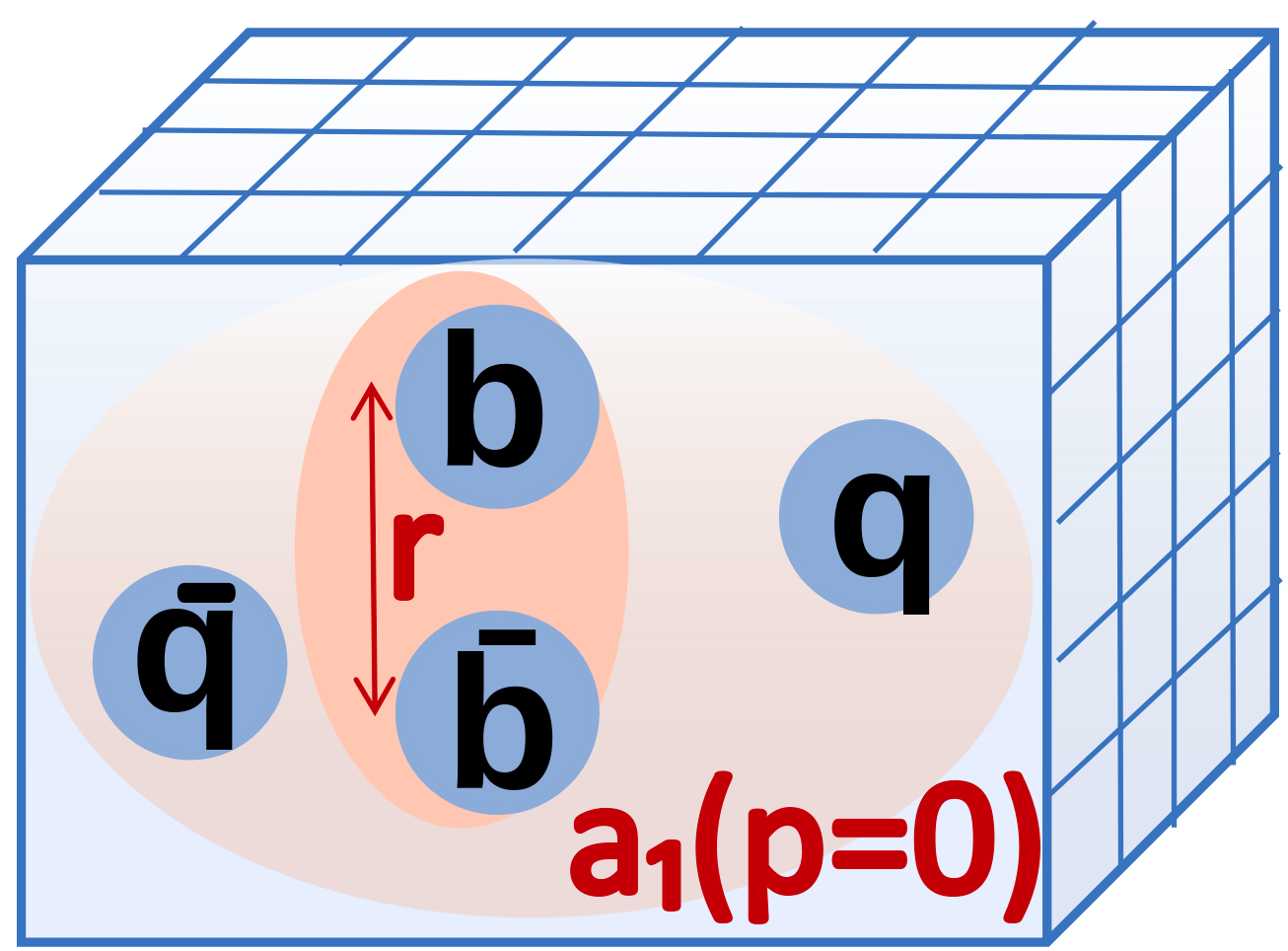}
	\end{subfigure}
	\vspace{-0.1cm}
							   &    
	\begin{subfigure}{0.18\textwidth}
	\includegraphics[width=\linewidth]{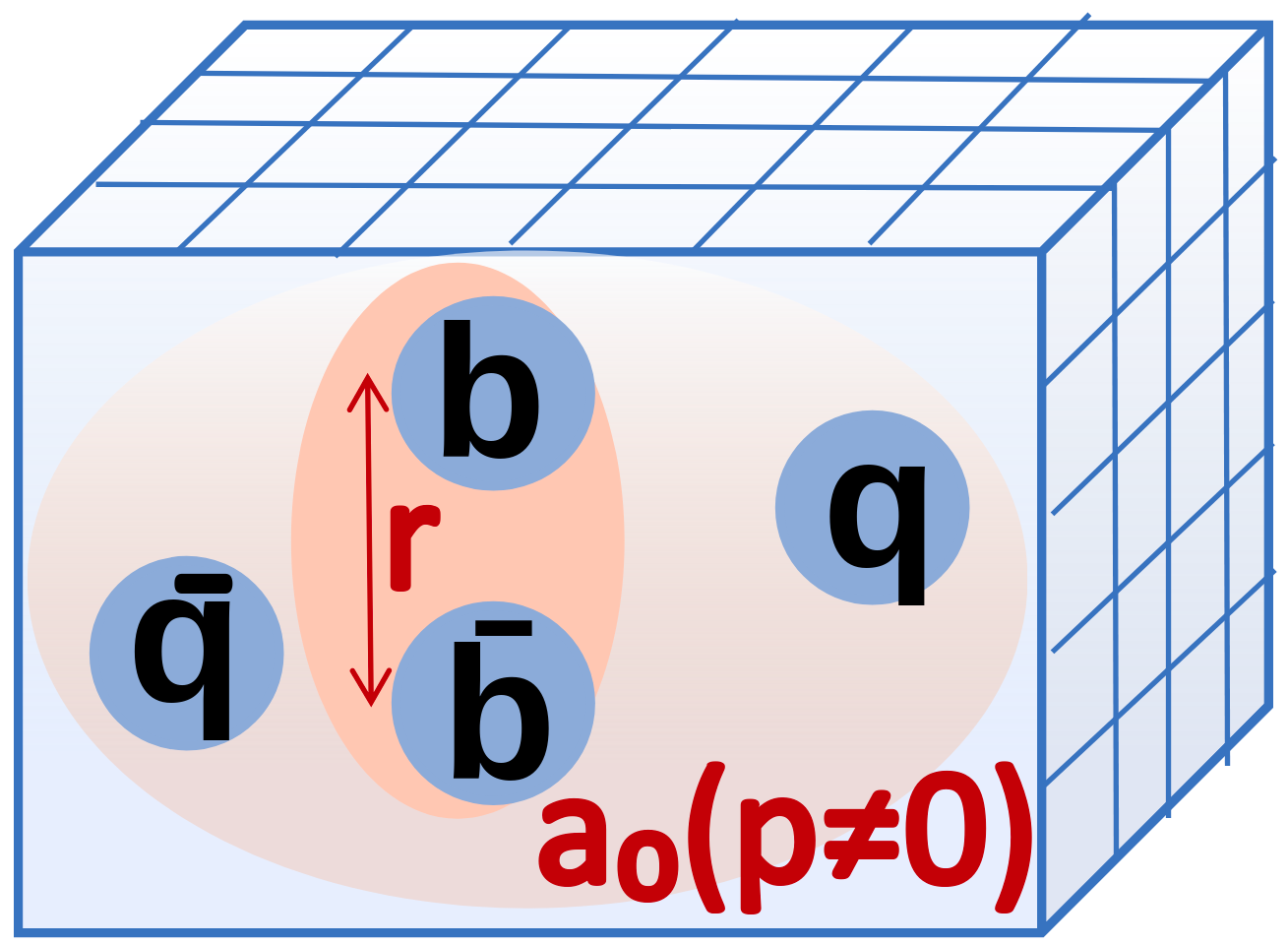}
	\end{subfigure}     
									   \\
                                     &   
	\begin{subfigure}{0.18\textwidth}
	\vspace{2.75cm}
	\caption{$J_z^l=0,\ C\!\cdot\! P=+1,$ $\epsilon=+1$} \label{fig:1b}	
	\end{subfigure}     
						  &   
	\begin{subfigure}{0.18\textwidth}
	\includegraphics[width=\linewidth]{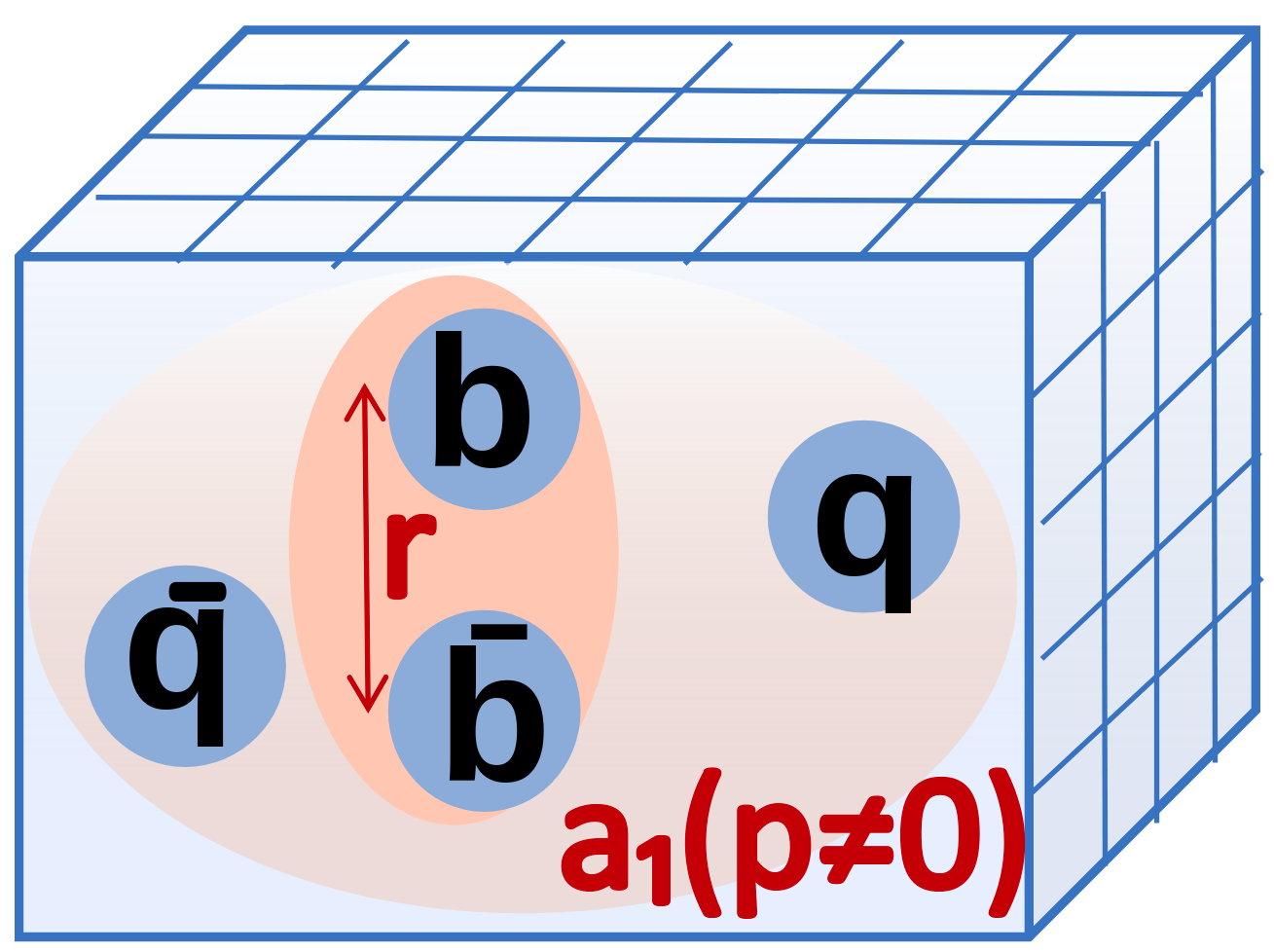}
	\caption{$J_z^l=0,\ C\!\cdot\! P=+1,$ $\epsilon=-1$} \label{fig:1c}	
	\end{subfigure}     
							   &    	
	\begin{subfigure}{0.18\textwidth}
	\vspace{2.75cm}
	\caption{$J_z^l=0,\ C\!\cdot\! P=-1,$ $\epsilon=+1$} \label{fig:1d}	
	\end{subfigure}     	   \\
\end{tabular}
\caption{(a) The system studied  with static $b$ and $\bar b$; (b,c,d) The states of this system   with various quantum numbers captured by the operators in (\ref{E3},\ref{E4},\ref{E5}).}\label{fig:1}
\end{figure*}

We perform a lattice QCD simulation of  the system $\bar bb\bar q_1q_2$  with isospin one, where $b$ and $\bar b$ quarks are static and fixed at distance $r$ (see Fig. \ref{fig:1a}). Systems containing $\bar q_1q_2\!\propto \!\bar du, \bar uu\!-\!\bar dd, \bar ud$ are equivalent in our simulation with $m_u\!=\!m_d$ that neglects the electromagnetic interaction. Therefore we  present the simulation for the neutral system $\bar bb\bar qq\!\propto \!\bar bb(\bar uu\!-\!\bar dd)$ where the charge conjugation for $\bar qq $ is a good quantum number. The goal is to determine the eigenenergies of this system $E_n(r)$ as a function of the separation $r$ for various quantum numbers. The resulting energies are then compared to the noninteracting (n.i.) energies $E^{\textrm{n.i.}}(r)$ of subsystems $[\bar bb][\bar qq]$ and $[\bar bq][\bar qb]$, where $[..]$ denotes a color-singlet meson of a given flavor. The eigenenergies  represent lattice input to study this system within the Born-Oppenheimer approximation via static potentials, according to the general strategy outlined in \cite{BO,Braaten:2014qka,Brambilla:2017uyf,Brambilla:2018pyn,Brambilla:2019jfi,Soto:2020xpm}. This  approximation is valuable when the $b$-quark mass is much larger than  the energy scale of the light degrees of freedom. The gluon and light-quark fields respond almost instantaneously to the motion of the $b$ and $\bar b$. Their instantaneous configurations are determined by the positions of the $b$ and $\bar b$, which are approximated as static color sources. The energy of stationary configurations of light-quark and gluon fields defines the static potential $V(r)$, which depends on the separation $r$ and the quantum numbers for the system.

Let us consider which quantum numbers of the system $\bar bb\bar qq$ are most relevant for the $Z_b$ resonances. The total spin of heavy quarks ($S^h$) and the angular momentum of the light degrees of freedom ($J^l$) are separately conserved in the static limit $m_b\!\to\! \infty$. The $Z_b$ with $J^P\!=\!1^+$ corresponds in the molecular $B^{(*)}\bar B^{*}$ picture to the linear combination of two quantum channels

\begin{align}
\label{decomposition}
B \bar B^*_k + B^*_k\bar B
&\propto (S^h\!=\!0)(J^l\!=\!1)+(S^h\!=\! 1)(J^l\! =\! 0)\>, \nonumber\\ 
B^*_i \bar B^*_j - B^*_j\bar B^*_i
& \propto (S^h\!=\!0)(J^l\!=\!1)- (S^h\!=\!1)(J^l\!=\!0)\>.
\end{align}
$Z_b$ within a diquark-antidiquark $(\bar b \bar q)_{3_c}(b q)_{\bar 3_c}$ picture is also a linear combination of these two quantum channels.

Lattice simulations of $Z_b$ \cite{Peters:2016wjm, Prelovsek:2019ywc} have been done only for the quantum number  $J^l\! =\! 0$, where $Z_b$ couples to $\Upsilon\;\pi$ and to the second component of  $B^{(*)}\bar B^*$ on the right-hand side of \eqref{decomposition} (\footnote{This is the quantum channel denoted by  $\Sigma^-_u$ in Table \ref{tab:1}.}). Throughout this paper we refer to any combination of $B^{(*)}\bar B^{(*)}$ as $B\bar B^{*}$ (\footnote{One should note that in the static limit $B$ and $B^*$ mesons are degenerate and with $B$, $B^*$ we refer to B-mesons with negative parity.}). Both available studies found that the eigenstate dominated by $B\bar{B}^*$ has energy significantly below $m_B+m_{B^*}$ at small $r$. This rendered the static potential with sizable attraction  between $B$ and $\bar B^*$ at small $r$. The Schr{\"o}dinger equation for $B\bar{B}^*$ leads to a bound state below the $B\bar{B}^*$ threshold, which could be related to $Z_b$.

The present  lattice study considers another three sets of quantum numbers for the $\bar bb\bar qq$ system. These quantum numbers have not been  studied before,  with exception of \cite{Alberti:2016dru}  which  considered  the ground state of one channel as detailed in Sec. \ref{sec:comparison}.  We investigate  the quantum number which contains $J^l\! =\!1$ and is relevant  for $Z_b$, where this resonance couples to $\eta_b\;\rho$ and to the first component of $B\bar B^*$ \eqref{decomposition}. In addition, we study two other sets of quantum numbers which do not couple to  $B\bar B^*$ but only to $[\bar bb][\bar qq]$ in the explored energy region.

\begin{table}[h!]
\centering
\begin{tabular}{ccccc|c|cc|c} 
\hline
\hline
\multicolumn{8}{c|}{Quantum numbers}                                                                                & \multirow{2}{*}{Lattice  studies}           \\ 
\cline{1-8}
 $I$                & $I_3$              & $J_z^{l}$ & $C\!\cdot\! P$ & $\epsilon$ & $\Lambda^{\epsilon}_{C\!P}$  & $S^h$ & $S^h_z$ &                             \\ 
\hline
 \multirow{4}{*}{\vspace{-0.3cm}1} & \multirow{4}{*}{\vspace{-0.3cm}0} & \multirow{4}{*}{\vspace{-0.3cm}0} & $-1$       & $-1$       & $\Sigma_u^-$       &  \multirow{4}{*}{\vspace{-0.3cm}0,1}  & \multirow{4}{*}{\vspace{-0.3cm}0}         & \cite{Prelovsek:2019ywc,Peters:2016wjm}                        \\ 
\cline{9-9}
                     &                    &                    & $+1$       & $+1$       & $\Sigma_g^+$   &&            & \multirow{3}{*}{This work}  \\
                   &                    &                    & $+1$       & $-1$       & $\Sigma_g^-$           &&    &                             \\
                   &                    &                    & $-1$       & $+1$       & $\Sigma_u^+$           &&    &                             \\
\hline
\hline
\end{tabular}
\caption{Four sets of quantum numbers for the system $\bar bb\bar qq$: the first one was studied in \cite{Prelovsek:2019ywc,Peters:2016wjm}, whereas we study the other three.   The system is invariant under the rotations of the heavy quark spins, so  the results are independent of $S^h$. $\Lambda^{\epsilon}_{\eta=C\!P}$ is written according to the convention in \cite{Juge:1999ie}.}
\label{tab:1}
\end{table}

\section{Quantum numbers and operators}\label{sec:QNandOP}

In the static approximation $m_b\to \infty$, the conserved quantum numbers differ from those when $b$ and $\bar{b}$ have finite mass.  Here are the quantum numbers  that characterize the $\bar{b}b\bar{q}q$ system in Fig. \ref{fig:1a} where  $b$ and $\bar b$ are separated along the $z$-axis:
\begin{itemize}[noitemsep,nolistsep] % you need \usepackage{enumitem}
\item $I\!=\!1$ and $I_3\!=\!0$: isospin    and its third component 
\item Angular momentum ($J^l_z=0$): The static quarks can not flip spin via interactions with gluons. Therefore the total spin of the heavy quarks ($S^h$) and the angular momentum of light degrees of freedom ($J^l$) are separately conserved.  The observables do not depend on $S^h$ due to the heavy quark symmetry, so the resulting eigenenergies apply to both $S^h\!=\!1$ and $S^h\!=\!0$. The only conserved rotational symmetry are the rotations around the separation axis $z$, so only the $z$-component of $J^l$ is  conserved.  
\item $C\!\cdot\! P=\pm 1$: The product of parity (space inversion with respect to the midpoint between $b$ and $\bar{b}$) and charge conjugation for light degrees of freedom is a good symmetry in the case of a neutral system $\bar bb\bar q q$.
\item $\epsilon=\pm 1$: This is an eigenvalue related to the reflection of the light degrees of freedom over $yz$ plane.  It is a good quantum number for  $J^{l}_z=0$.  
\end{itemize}

%\vspace{0.2cm}

We  study    the   four-quark system $\bar{b}b\bar{q}q$   with  
 \begin{equation}
\label{E2}
I=1,\ I_3=0, \  J^{l}_z=0~, 
\end{equation}
where the  operators are written for   $I_3=0$ due to considerations related to the charge conjugation.
Table \ref{tab:1} lists the  three sets of quantum numbers considered here and one set considered  in the previous studies \cite{Peters:2016wjm, Prelovsek:2019ywc}.
To show the the connection between the $Z_b$ in the molecular picture and the quantum channels we consider, let us write Eq. \eqref{decomposition} more rigorously using Fierz transformations,

\begin{widetext}
\begin{align}
\label{decomposition-detailed}
B \bar B^*_k + B^*_k\bar B &\propto  (S^h\!=\!0)(J^l\!=\!1,\  C\! \cdot\! P\!=\!\epsilon\!=\!+1)+(S^h\!=\! 1)(J^l\! =\! 0,\  C\!\cdot\!P\!=\!\epsilon\!=\!-1)\nonumber\\
[\bar b P_- \gamma_5 q] ~[\bar q \gamma_z P_+ b] + [\bar b P_- \gamma_z q] ~[\bar q \gamma_5 P_+ b]&=
\left(\bar b^a\gamma_5 P_+b^b\right)\left(\bar q^bP_-\gamma_{z}q^a\right) + \left( \bar b^a\gamma_zP_+b^b\right)\left(\bar q^bP_-\gamma_{5}q^a\right)\>, \nonumber\\[7pt]
B^*_i \bar B^*_j - B^*_j\bar B^*_i & \propto  (S^h\!=\!0)(J^l\!=\!1,\  C\! \cdot\! P\!=\!\epsilon\!=\!+1)- (S^h\!=\! 1)(J^l\! =\! 0,\  C\!\cdot\!P\!=\!\epsilon\!=\!-1)\nonumber\\
[\bar b P_- \gamma_x q] ~[\bar q \gamma_y P_+ b] - [\bar b P_- \gamma_y q] ~[\bar q \gamma_xP_+ b]&=
\left(\bar b^a\gamma_5 P_+b^b\right)\left(\bar q^bP_-\gamma_{z}q^a\right) - \left( \bar b^a\gamma_zP_+b^b\right)\left(\bar q^bP_-\gamma_{5}q^a\right) \>, 
\end{align}
\end{widetext}
where color indices $a$, $b$ are summed over and $P_{\pm}=(\mathbb{1}\pm\gamma_t)/2$. The right-hand side is a linear combination of two terms; the first one is the quantum channel we consider and the second one was considered in \cite{Peters:2016wjm, Prelovsek:2019ywc}. The $Z_b$ in diquark-antidiquark picture  $\left(\bar b P_- C \gamma_5 \bar q^T\right)_{3_c}\!\!\left(b^T P_+\gamma_z C q\right)_{\bar 3_c}~\!\!\!\!-~\!\!\left(\bar b P_- C \gamma_z \bar q^T\right)_{3_c}\!\!\left(b^T P_+\gamma_5 C q\right)_{\bar 3_c}$ is also a linear combination of these two quantum channels.

We determine the eigenenergies $E_n$ of the system in Fig. \ref{fig:1a} from the correlation functions $\langle O_i (t)O_j^\dagger(0)\rangle$. Our operators  resemble Fock components $[\bar b q][\bar qb]$ and $[\bar b b][\bar qq]$, schematically shown in Fig. \ref{fig:1} with quantum numbers represented in Table \ref{tab:1}. Employed annihilation operators  for each set of quantum numbers are listed below, followed by comments on the notation and various constructions,
{\small
\begin{align}
\label{E3}
 & J_z^l=0,\ C\!\cdot\! P= +1,\ \epsilon=+1 \ , \ S^h_z=0,\  S^h=0~\mathrm{or} ~1\>, ~ \nonumber\\[5pt]
O_1\!&=\! O_{B\bar B^*}\!\!\propto\sum_{a,b}\sum_{A,B,C,D}\!\!\!\!\Gamma_{BA}\tilde \Gamma_{CD}~\bar b^a_C(0)q_A^a(0)~ \bar q^b_B(r)b_D^b(r) \nonumber \\ 
% each of the 4 terms in a new row + correction
&\propto \left[\bar b(0) P_- \gamma_5 q(0)\right]~\left[\bar q(r) \gamma_z P_+ b(r)\right] \nonumber \\ 
&+ \left[\bar b(0) P_- \gamma_z q(0)\right]~\left[\bar q(r) \gamma_5 P_+ b(r)\right]   \nonumber \\ 
& -  \left[\bar b(0) P_- \gamma_y q(0)\right]~\left[\bar q(r) \gamma_x P_+ b(r)\right] \nonumber \\ 
&+  \left[\bar b(0) P_- \gamma_x q(0)\right]~\left[\bar q(r) \gamma_y P_+ b(r)\right]  \>,  \nonumber \\[5pt]
O_2\!&=\! O_{(B\bar B^*)'}\>, \nonumber\\[5pt]
O_3\!&=\!O_{[\bar{b}b] \rho(0)}\!\propto\! [\bar b(0) U\Gamma^{\textrm{(H)}} b(r)]~[\bar q \gamma_zq]_{\vec p=\vec 0}\>,\nonumber\\[5pt]
O_4\!&=\!O_{[\bar{b}b] \rho(1)}\!\propto\! [\bar b(0) U\Gamma^{\textrm{(H)}} b(r)]~ \bigl([\bar q \gamma_zq]_{\vec p=\vec e_z}+ [\bar q \gamma_zq]_{\vec p=-\vec e_z}\bigr)\>,\nonumber\\[5pt]
O_5\!&=\!O_{[\bar{b}b] \rho(2)}\!\propto\! [\bar b(0) U\Gamma^{\textrm{(H)}} b(r)]~ \bigl([\bar q \gamma_zq]_{\vec p=2\vec e_z}+ [\bar q \gamma_zq]_{\vec p=-2\vec e_z}\bigr) \>,
\end{align}
} 

{\small
\begin{align}
\label{E4}
 &\   J_z^l=0,\ C\!\cdot \! P= +1,\ \epsilon=-1\ , \ S^h_z=0,\  S^h=0~\mathrm{or} ~1\>,\nonumber\\[5pt]
O_1\!&=\!O_{[\bar{b}b] \pi(1)}\!\propto\! [\bar b(0) U\Gamma^{\textrm{(H)}} b(r)]~\bigl([\bar q \gamma_5q]_{\vec p=\vec e_z}- [\bar q \gamma_5q]_{\vec p=-\vec e_z}\bigr)\>,\nonumber\\[5pt]
O_2\!&=\!O_{[\bar{b}b] \pi(2)}\!\propto\! [\bar b(0) U\Gamma^{\textrm{(H)}} b(r)]~\bigl([\bar q \gamma_5q]_{\vec p=2\vec e_z}- [\bar q \gamma_5q]_{\vec p=-2\vec e_z}\bigr)\>,\nonumber\\[5pt]
O_3\!&=\!O_{[\bar{b}b] b_1(1)}\!\propto\! [\bar b(0) U\Gamma^{\textrm{(H)}} b(r)]~\bigl([\bar q \gamma_x\gamma_yq]_{\vec p=\vec e_z}- [\bar q \gamma_x\gamma_yq]_{\vec p=-\vec e_z}\bigr)\>,\nonumber\\[5pt]
O_4\!&=\!O_{[\bar{b}b] a_1(0)}\!\propto\! [\bar b(0) U\Gamma^{\textrm{(H)}} b(r)]~ [\bar q \gamma_5\gamma_zq]_{\vec p=\vec 0}\>,\nonumber\\[5pt]
O_5\!&=\!O_{[\bar{b}b] a_1(1)}\!\propto\! [\bar b(0) U\Gamma^{\textrm{(H)}} b(r)]~ \bigl([\bar q \gamma_5\gamma_zq]_{\vec p=\vec e_z}+ [\bar q \gamma_5\gamma_zq]_{\vec p=-\vec e_z}\bigr)\>,
\end{align}
}
 {\small
\begin{align}
\label{E5}
 &\   J_z^l=0,\ C\!\cdot\! P= -1,\ \epsilon=+1\ , \ S^h_z=0,\  S^h=0~\mathrm{or} ~1\>,\nonumber\\[5pt]
O_1\!&=\!O_{[\bar{b}b] \rho(1)}\!\propto\! [\bar b(0) U\Gamma^{\textrm{(H)}} b(r)]~\bigl([\bar q \gamma_zq]_{\vec p=\vec e_z}- [\bar q \gamma_zq]_{\vec p=-\vec e_z}\bigr)\>,\nonumber\\[5pt]
O_2\!&=\!O_{[\bar{b}b] \rho(2)}\!\propto\! [\bar b(0) U\Gamma^{\textrm{(H)}} b(r)]~\bigl([\bar q \gamma_zq]_{\vec p=2\vec e_z}- [\bar q \gamma_zq]_{\vec p=-2\vec e_z}\bigr)\>,\nonumber\\[5pt]
O_3\!&=\!O_{[\bar{b}b] a_0(1)}\!\propto\! [\bar b(0) U\Gamma^{\textrm{(H)}} b(r)]~\bigl([\bar q \mathbb{1}q]_{\vec p=\vec e_z}- [\bar q \mathbb{1} q]_{\vec p=-\vec e_z}\bigr)\>.
\end{align}
}

Let us first provide general comments on all operators, followed by comments specific to both operator types. Color singlets are denoted by $[..]$. The gamma matrices sandwiched between static quarks  are $\tilde\Gamma,\Gamma^{\textrm{(H)}}=\gamma_5P_+$  or $ \gamma_zP_+$ for $S^h=0$ or $1$, respectively. The heavy quark symmetry implies that the correlators and  $E_n$ are the same for both, so our results apply to both cases. The pair $\bar qq$  indicates the combination $\bar uu-\bar dd$ with $I\!=\!1$ and $I_3\!=\!0$.  All light quarks $q(x)$ are smeared around the position $x$ using the full distillation \cite{Peardon:2009gh} of a radius about \SI{0.3}{fm}, while the heavy quarks are pointlike. Additional arguments concerning quantum numbers of analogous operators are given in the appendix of \cite{Prelovsek:2019ywc}.

The operators $O_{B\bar B^*}$  resembling $[\bar bq][\bar qb]$  are constructed with $\Gamma\!\!=\!\!P_-\gamma_z$ that satisfies   $J^{l}_z=0$. The small and capital letters denote color and Dirac indices, respectively.  The explicitly written four terms of $O_1$ in \eqref{E3} are obtained via the Fierz transformation, where we take $\tilde \Gamma\!=\!\gamma_5 P_+$. This decomposition clarifies that this quantum channel is a linear superposition of $B\bar B^*$, $B^* \bar B$, and $B^* \bar B^*$. For all three sets of quantum numbers more operators of this kind could be constructed, but we will limit ourselves to those where both $B$-mesons have negative parity, since they correspond to the ground states\footnote{Only operators where exactly one $B$-meson carries positive parity could contribute to last two quantum channels.}. $O_{(B\bar B^*)'}$ is obtained from $O_{B\bar B^*}$ by replacing all $q(x)$ with $\nabla^2 q(x)$.

The operators resembling $[\bar bb][\bar qq]$ are formed from a color-singlet bottomonium and color-singlet light-meson current. The static quarks are  connected by the product of gauge-links $U$  between $0$ and $r$. The light degrees of freedom $[\bar q \Gamma^\prime q]_{\vec p}\equiv \tfrac{1}{V} \sum_{\vec x} \bar q(\vec x) \Gamma^\prime q(\vec x) e^{i \vec p\vec x}$ with $I\!=\!1$ are projected to definite momenta $\vec p=\vec n \tfrac{2\pi}{L}$ (given in units of $2\pi/L$), which can be different since momentum is not conserved in the presence of static quarks. The signs between the two terms ensure the right $C\!\cdot\! P$ and the momenta are in the $z$-direction due to $J^{l}_z=0$. The light current  $[\bar q \Gamma^\prime q]$ with $\Gamma^\prime=\gamma_5$ couples in the low-energy region to a pion, $l=\pi$. The currents for other $\Gamma^\prime$ couple to resonances $l=\rho,~b_1~a_1,~a_0$ that are not strongly stable on our lattice. So, these currents in principle also couple to the allowed strong decay products $\pi\pi$, $\omega\pi,$ $\pi\pi\pi$, and $\pi\eta^\prime$, respectively (we listed just few examples relevant for the simulation with $N_{\textrm{f}}=2$). The reliable and rigorous extraction of eigenenergies would require implementation of multihadron operators in the light sector which is beyond the scope of the present study. In practice, the employed operator $[\bar q \Gamma^\prime q]_{\vec p}$ couples to one finite-volume energy level with energy $E_{l(\vec p)}$ in the low-energy region. We refer to this level  as $l=\rho,~b_1,~a_1,~a_0$; however, this level is a mixture of resonant and multihadron eigenstates in practice. Our main purpose is  to find out whether there is some interaction between bottomonium $[\bar bb]$ and the light degrees of freedom  $l$ for given quantum numbers. Therefore we will compare the sum of the separate energies $V_{\bar bb}+E_{l(\vec p)}$ with the eigenenergy $E_n$ of the whole system $[\bar bb][\bar qq]$, where the light degrees of freedom arise from the same current $[\bar q \Gamma^\prime q]_{\vec p}$ in both cases.  This strategy will not lead to the complete spectrum of eigenenergies, but it will still indicate whether the energy of the light degrees of freedom is affected in the presence of the bottomonium.
 
This choice of operators captures all states with noninteracting energy  below   $2m_B\pm\SI{50}{MeV}$. This holds for all three quantum channels with a subtlety for $C\!\cdot\! P\!=\!\epsilon\!=\!+1$, where operators $[\bar bb] a_0(p\!=\!1,2)$ are implemented but not used in the analysis due to the noisy plateaus.
 
\begin{table}[h!]
\centering
\scalebox{0.92}{
\begin{tabular}{c|llllll} 
\hline
\hline
$J_z^{l}$ &\multicolumn{2}{c}{0}          & \multicolumn{2}{c}{0}          & \multicolumn{2}{c}{0}             \\ 
$C\!\!\cdot\!\! P$ &\multicolumn{2}{c}{$+1$}          & \multicolumn{2}{c}{$+1$}          & \multicolumn{2}{c}{$-1$}             \\ 
$\epsilon$ &\multicolumn{2}{c}{$+1$}          & \multicolumn{2}{c}{$-1$}          & \multicolumn{2}{c}{$+1$}             \\ 
\hline
&\multicolumn{2}{l}{}                                          & \multirow{5}{*}{$[\bar b(0) b(r)]$} & $\pi(\vec{p}=\vec{e}_z)$  &                                   &                             \\
&\multicolumn{2}{c}{$B(0) \bar B^*(r)$}                        &                                   & $\pi(\vec{p}=2\vec{e}_z)$ & \multirow{3}{*}{$[\bar b(0) b(r)]$} & $\rho(\vec{p}=\vec{e}_z)$   \\
&\multirow{3}{*}{$[\bar b(0) b(r)]$} & $\rho(\vec{p}=\vec{0})$    &                                   & $b_1(\vec{p}=\vec{e}_z)$  &                                   & $\rho(\vec{p}=2\vec{e}_z)$  \\
&                                  & $\rho(\vec{p}=\vec{e}_z)$  &                                   & $a_1(\vec{p}=\vec{0})$    &                                   & $a_0(\vec{p}=\vec{e}_z)$    \\
&                                  & $\rho(\vec{p}=2\vec{e}_z)$ &                                   & $a_1(\vec{p}=\vec{e}_z)$  &                                   &                             \\
\hline
\hline
\end{tabular}}
\caption{The relevant states of the $\bar{b}b\bar{q}q$ system with $I\!=\!1$ captured by the operators in \eqref{E3}, \eqref{E4}, and \eqref{E5}, respectively.}
\label{tab:2}
\end{table}

\section{Lattice details}\label{sec:details}

Simulation is performed on an ensemble with dynamical Wilson-clover $u/d$ quarks,  $m_\pi \simeq \SI{266(5)}{MeV}$, $a\simeq \SI{0.1239(13)}{fm}$ and 281 configurations \cite{Hasenfratz:2008ce,Lang:2011mn}. We employ an ensemble with small $N_L\!=\!16$ and $L\!\simeq\! \SI{2}{fm}$, since larger $L$ would require more operators with higher $\vec p$ to study the same energy region. The lattice temporal extent $N_T\!=\!32$ is effectively  doubled by summing the light-quark propagators with periodic and antiperiodic boundary conditions in time \cite{Lang:2011mn}.

\section{Calculation of eigenenergies and overlaps}\label{sec:overlaps}

For the evaluation of correlation matrices $C_{ij}(t)=\langle O_i (t)O_j^\dagger(0)\rangle$, the full distillation method  \cite{Peardon:2009gh} is used. The dimensions of $C_{ij}$ with operators \eqref{E3}, \eqref{E4} and \eqref{E5} is $5\times5$, $5\times5$, and $3\times3$, respectively. The $\bar bb$ annihilation Wick contraction is not present in the static limit considered here. $C_{ij}$ that contain $O_{B\bar B^*}$, $O_{(B\bar B^*)'}$ are averaged over $8^3$ space coordinate starting points of $\bar b$, while all other matrix elements are averaged over $16^3$ space positions, over all source time slices, over all three possible directions of momenta, and all three possible polarizations of gamma matrices. We extract eigenenergies $E_n$ and overlaps $\langle O_i|n\rangle$ from the matrices $C_{ij}(t)=\sum_n \langle O_i|n\rangle e^{-E_nt} \langle n|O_j^\dagger\rangle$ using the generalized eigenvalue problem (GEVP) \cite{Michael:1985ne,Luscher:1990ck,Blossier:2009kd} and $t_0/a=2$. 

\section{Eigenenergies of  $\bar bb\bar qq$ system with $I\!=\!1$ as a function of $r$}\label{sec:eigen-energies}

\begin{figure}[h!]
\begin{center}
\includegraphics[width=0.47\textwidth]{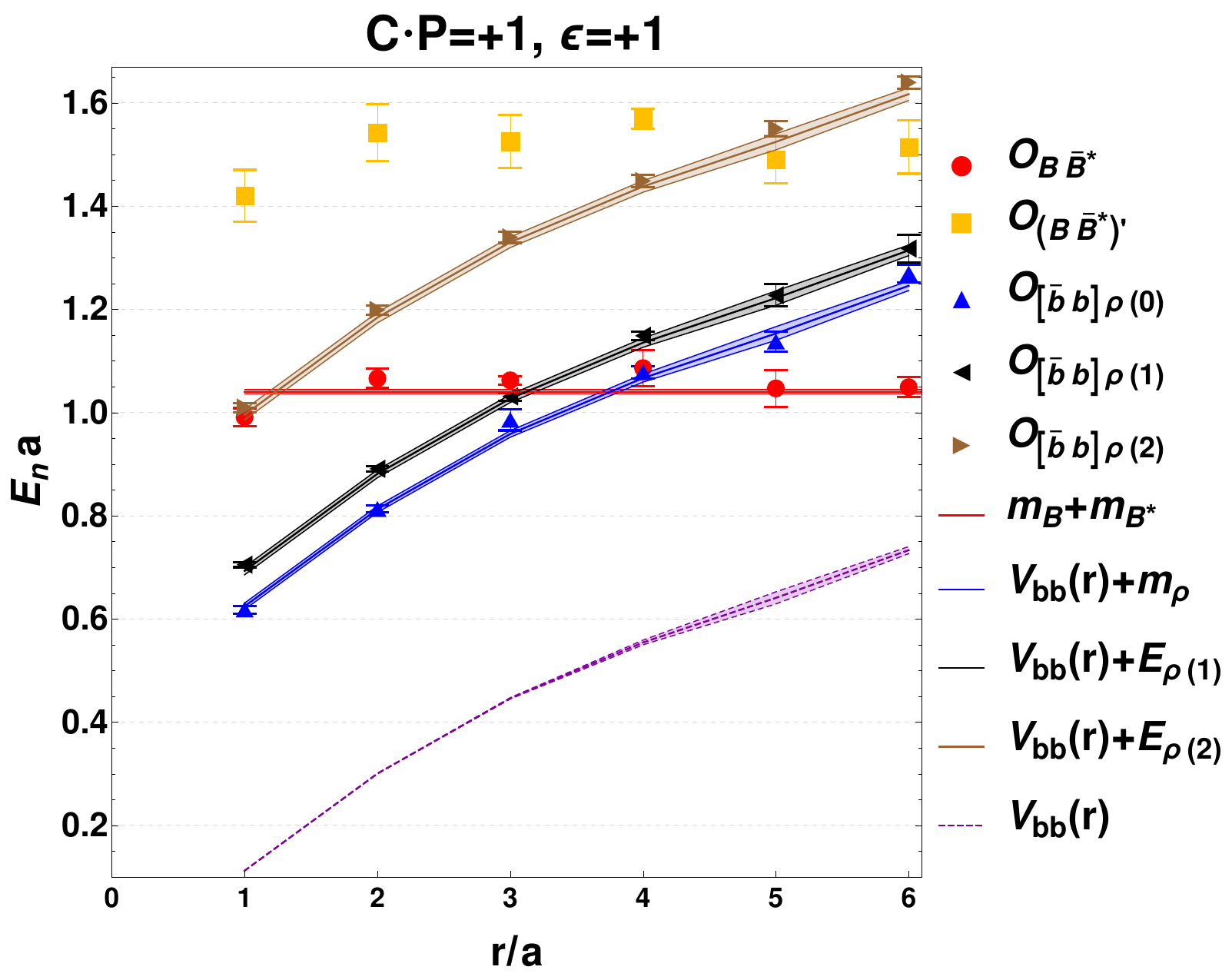}
\caption{Results for the system $\bar bb\bar qq$ with  quantum numbers $I\!=\!1,~J^l_z\!=\!0,~C\!\cdot\! P\!=\!+1, ~ \epsilon\!=\!+1$.  Eigenenergies  are shown by symbols  for separations between  static quarks $b$ and $\bar b$ up to $r=6a$. The labels indicate which two-hadron component dominates each eigenstate. The lines represent related two-hadron energies $E^{\textrm{n.\,i.}}$ (\ref{E6}) when two hadrons (see Table \ref{tab:2}) do not interact. The width of their bands shows the uncertainty. The violet dashed line represents the static potential $V_{\bar bb}(r)$ between $b$ and $\bar b$. Lattice spacing is $a\simeq\SI{0.124}{fm}$.}
\label{fig:2}
\end{center}
\end{figure}

\begin{figure}[h!]
\begin{center}
\includegraphics[width=0.47\textwidth]{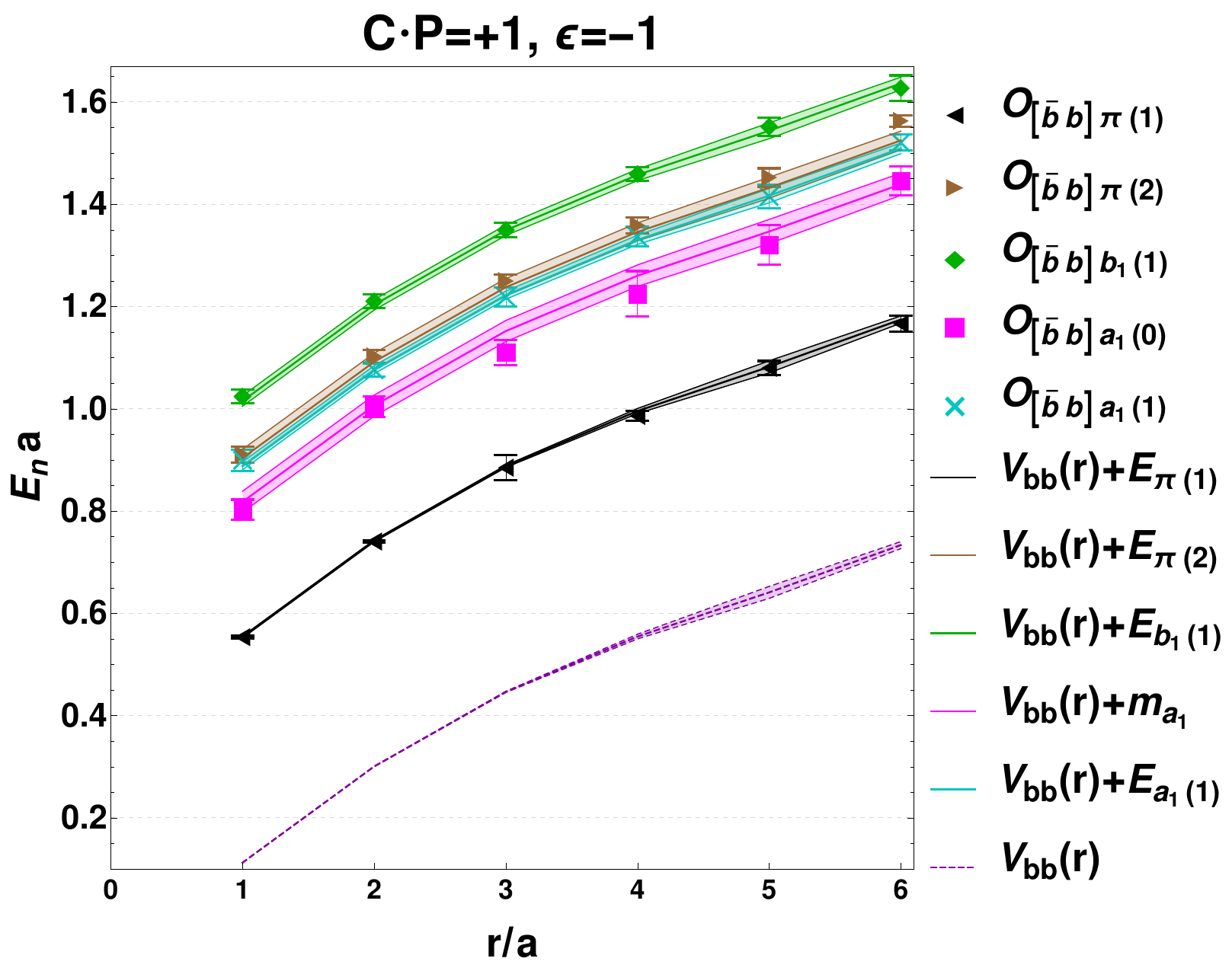}   
\caption{This figure shows the eigenenergies $E_n$ (symbols) and the two-hadron noninteracting energies $E^{\textrm{n.\,i.}}$ (lines) similarly as in Fig. \ref{fig:2}, but  for quantum numbers $I\!=\!1,~J^l_z\!=\!0,~C\!\cdot\! P=\!+1, ~ \epsilon\!=\!-1$.}
\label{fig:3}
\end{center}
\end{figure}

\begin{figure}[h!]
\begin{center}
\includegraphics[width=0.47\textwidth]{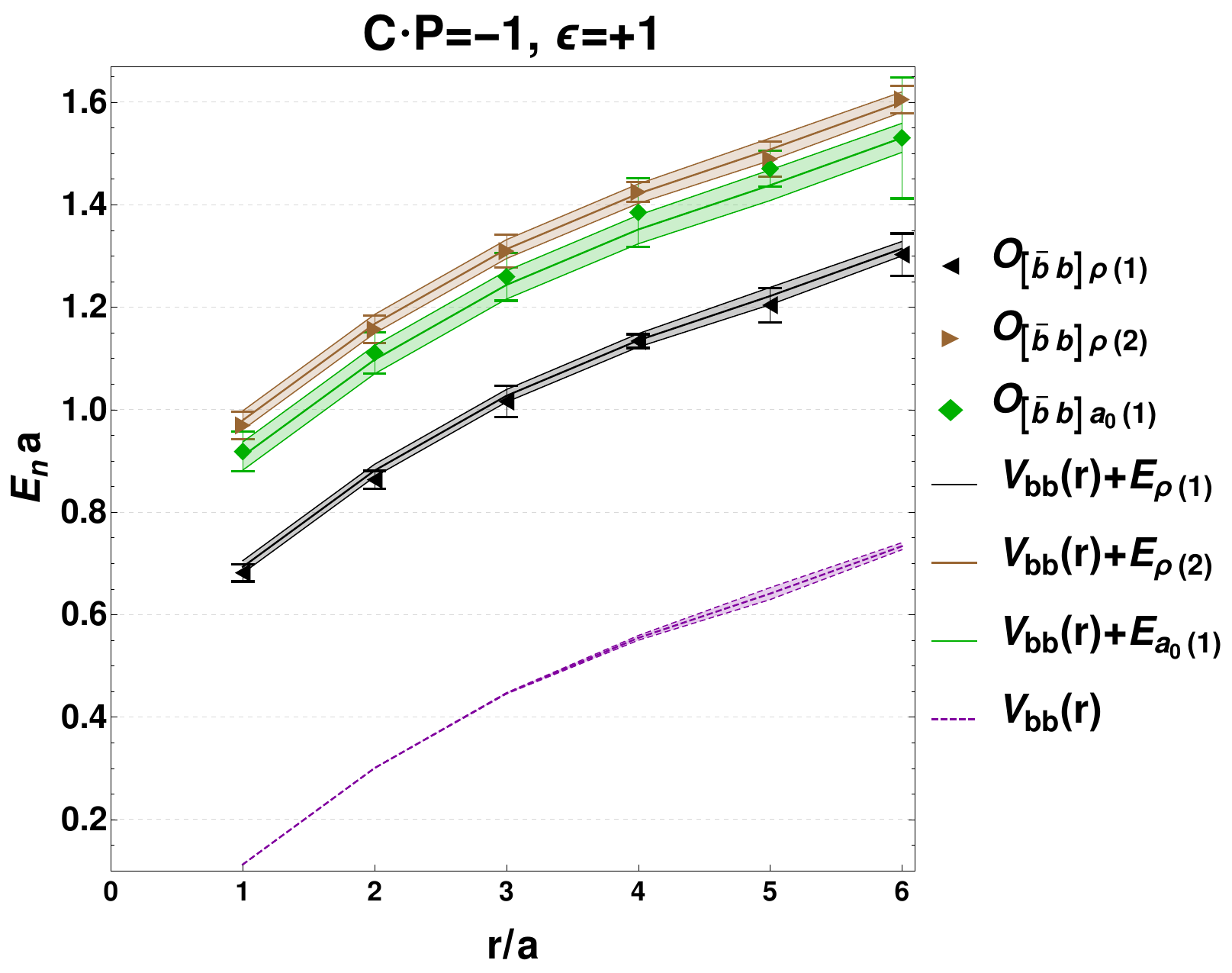}   
\caption{This figure shows the eigenenergies $E_n$ (symbols) and the two-hadron noninteracting energies $E^{\textrm{n.\,i.}}$ (lines) similarly as in Fig. \ref{fig:2}, but for quantum numbers $I\!=\!1,~J^l_z\!=\!0,~C\!\cdot\! P=\!-1, ~ \epsilon\!=\!+1$.}
\label{fig:4}
\end{center}
\end{figure}

The central results of our study are the eigenenergies of the $\bar bb\bar qq$ system with $I\!\!=\!\!1$ (Fig. \ref{fig:1a}) with static $b$ and $\bar b$ separated by $r$. Eigenenergies are presented by symbols in Figs. \ref{fig:2}-\ref{fig:4} for the three sets of quantum numbers shown in Table \ref{tab:1}. The colors of symbols indicate which Fock component (see Table \ref{tab:2}) dominates an eigenstate, as determined from the normalized overlaps  of an eigenstate $|n\rangle$ to operators $O_i$. The normalized overlap $\tilde Z_i^n\equiv \langle O_i|n\rangle/\max_m \langle O_i|m\rangle$ is normalized so that its maximal value for a given $O_i$ across all eigenstates is equal to one.  

The lines in Figs. \ref{fig:2}-\ref{fig:4} provide the related noninteracting (n. i.) energies $E_n$ of two-hadron states shown in Table \ref{tab:2}
\begin{equation}
\begin{split}
  E^{\textrm{n.\,i.}}_{B\bar B^*}\!&=\!2m_B,  \ E^{\textrm{n.\,i.}}_{[\bar b(0) b(r)] l(0)}\!=\!V_{\bar bb}(r)+m_{l}\>,\\    
  E^{\textrm{n.\,i.}}_{[\bar b(0) b(r)]l(\vec p)}\!&=\!V_{\bar bb}(r)+E_{l(\vec p)},\ l=\pi, \rho, b_1, a_1, a_0\>,
\label{E6}
\end{split}
\end{equation}
where $\bar bb$ static potential $V_{\bar bb}(r)$,   $m_{l}$ and $m_B=m_{B^*}=0.5201(19)$ (mass of $B^{(*)}$ for $m_b\to \infty$ without $b$ rest mass) are determined on the same lattice. The energy $E_{l(\vec p)}$ denotes the measured finite-volume energy that arises from the light current $[\bar q\Gamma^\prime q]_{\vec p}$, which turns out to be  $E_{l(\vec p)}\simeq \sqrt{m_l^2+\vec p^2}$ although some of the light hadrons $l$ are resonances.
   
All observed eigenenergies $E_n$ of the $\bar bb\bar qq$ system (symbols) are very close to noninteracting energies $E^{n.i.}$ of $[\bar bb][\bar qq]$ or $[\bar bq][\bar qb]$ (lines). This represents the most important conclusion of the present study. In particular, eigenstates dominated by $[\bar bb][\bar qq]$ operators have energies consistent with the sum of energies for $[\bar bb(r)]$ and $[\bar qq]$. Given our precision, we therefore do not observe attraction or repulsion between bottomonium and light hadrons for the considered separations $r$ between $b$ and $\bar b$. The absence of a sizable interaction is expected, since they do not share any valence quarks.  

    \begin{figure*}[ht!]
\centering
\begin{subfigure}{0.37\textwidth}
\includegraphics[width=\textwidth]{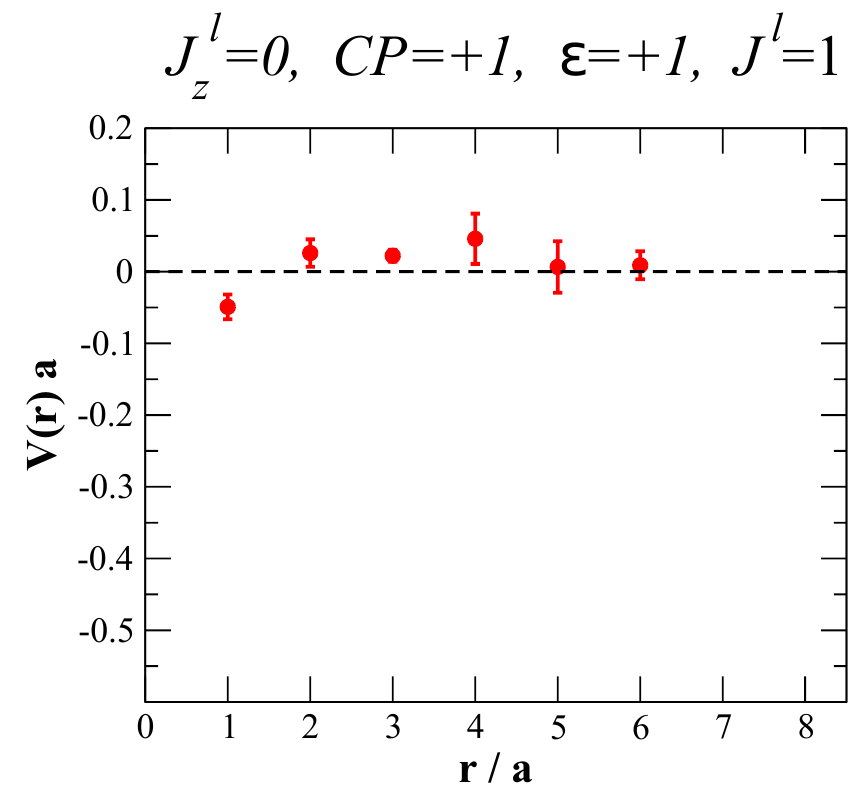}
\caption{ } \label{fig:5a}
\end{subfigure} 
\hspace*{1cm}
\begin{subfigure}{0.37\textwidth}
\includegraphics[width=\textwidth]{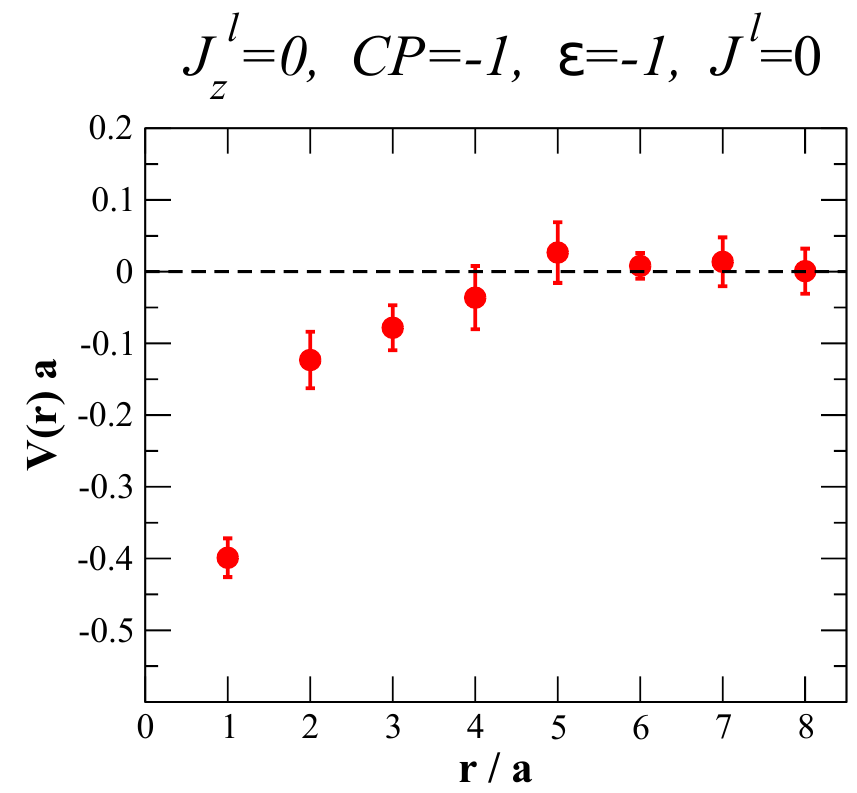}
\caption{ } \label{fig:5b}
\end{subfigure}
\caption{Static potentials between $B$ and $\bar  B^{*}$ separated by $r$ from lattice simulations (see Fig. \ref{fig:1a}).  Quantum numbers $\mathrm{(a)}\  I\!=\!1\!,~J^l\!=\!1,~J^l_z\!=\!0,~C\!\cdot\! P\!=\! \epsilon\!=\!+1$ are considered here and (b)  $I\!=\!1,~J^l\!=\!0,~J^l_z\!=\!0,~C\!\cdot\! P\!=\! \epsilon\!=\!-1$ were studied in    \cite{Prelovsek:2019ywc}. The potential (a) is consistent with zero for $r/a\geq 2$ within slightly more than one sigma errors, which are shown in the plot. Both simulations are performed on the same ensemble with the lattice spacing   $a\simeq\SI{0.124}{fm}$.}
\label{fig:5}
\end{figure*}

The eigenstate dominated by $B\bar B^*$ is present only for the quantum number $C\!\cdot\! P\!=\!\epsilon\!=\!+1$ that couples to $Z_b$. Its energy $E_{B\bar B^*}(r)$ is represented by the red circles in Fig. \ref{fig:2} and is close to $m_B+m_{B^*}$. This suggests that the interaction between $B$ and $\bar B^*$ is small in this quantum channel. For the remainder of the discussion, we assume that this eigenstate couples only to the $B\bar B^*$ Fock component and does not contain other Fock components, which is supported by the extracted normalized overlaps shown in Appendix \ref{sec:appendixA}. The energy of this eigenstate represents the total energy without the kinetic energy of heavy degrees of freedom. The difference $V (r) = E_{B\bar B^*}(r)-m_B-m_{B^*}$ therefore represents the potential felt by the heavy degrees of freedom, in this case between $B$ and $\bar B^*$ mesons. The extracted potential is shown in Fig. \ref{fig:5a}. It is consistent with zero for  $r\geq\SI{0.2}{fm}$ ($r/a\geq 2$), which implies that we do not find any significant attraction or repulsion between $B$ and $\bar B^*$ for these $r$ (\footnote{The potential for $r/a\geq 2$ is consistent with zero within slightly more than $1\sigma$.}).  The potential is slightly negative and therefore attractive at  $r\!=\! a\!\simeq\!  \SI{0.12}{fm}$ (\footnote{ This slightly attractive potential   agrees also with the result $-0.054\pm0.017$ based on the ratio method in Appendix \ref{sec:ratio}.})
   \begin{equation}
   \label{attraction}
   V(r\!=\!a)a=\left(E_{B\bar B^*}(r\!=\!a)-m_B-m_{B^*}\right)a=-0.049\pm 0.017\>,
   \end{equation}
which could hint at a  small attraction between $B$ and $\bar B^*$ at small $r$.  
Possible implications for $Z_b$ are discussed below. All these conclusions were verified also  considering the so-called ratio  method, discussed in Appendix \ref{sec:ratio}.

The $Z_b$ resonance is a linear superposition of two quantum channels listed in \eqref{decomposition-detailed} within the molecular picture; these two channels have different quantum numbers for light degrees of freedom, so their properties are not related by heavy quark spin symmetry:  
\begin{itemize}
\item   $J^l\!=\! 0~ \&~ C\!\cdot\! P\!=\! \epsilon\!=\!-1$: The potential with sizable attraction between $B$ and $\bar B^{*}$ at small $r$ has been found  \cite{Prelovsek:2019ywc,Peters:2016wjm}. The result from \cite{Prelovsek:2019ywc}, which is obtained on the same  ensemble as employed here, is shown in Fig. \ref{fig:5b}. This potential has been also obtained assuming that the eigenstate dominated by $B\bar B^*$ does not contain other Fock components. The motion of $B$ and $\bar B^*$ with experimental masses in this potential leads to one $B\bar B^*$ bound state below the threshold, whose binding energy depends on the parametrization of the potential; assuming the nonsingular potential $V(r)=-A r^{-(r/d)^F}$ leads to the range of binding energies $M-m_B-m_{B^*}=-48 ^{+41}_{-{108}}\,$MeV \cite{Prelovsek:2019ywc}. Some parametrizations among those lead to a bound state closely below threshold ($\simeq \SI{20}{MeV}$) and sharp peak in the $B\bar B^*$ rate above threshold $-$ a feature that could be related to the observed experimental $Z_b$ peak. Most of the parametrizations in \cite{Prelovsek:2019ywc} lead to a binding energy larger than $\SI{20}{MeV}$ and a less significant peak in the rate above the threshold, since the size of the peak decreases as the binding energy increases. The singular form of the potential $V(r)=-\tfrac{A}{r} r^{-(r/d)^F}$ would also  lead  to one bound state, but with a  larger binding energy\footnote{This has not been considered in \cite{Prelovsek:2019ywc}.}. This component is therefore significantly attractive; it is possible that this component alone is too attractive and leads to a binding energy that is too large in comparison with the experimental $Z_b$. 
\item $ J^l=1~ \&~ C\!\cdot\! P\!=\! \epsilon\!=\!+1$: The potential for this component in Fig. \ref{fig:5a} shows no observable attraction or repulsion between $B$ and $\bar B^{*}$ at $r\geq 0.2~$ and a very mild attraction at $r\simeq \SI{0.1}{fm}$ (\ref{attraction}).   
\item Linear combination: The $Z_b$ is a linear combination of those two quantum numbers \eqref{decomposition-detailed}. The $B\bar B^*$ and $ B^*\bar B^*$ channels are coupled in this system  via the  strongly attractive potential  for component  $C\!\cdot\! P\!=\! \epsilon\!=\!-1$ and very mildly attractive potential for $C\!\cdot\! P\!=\! \epsilon\!=\!+1$, both shown in Fig. \ref{fig:5}.  It is not possible to  establish implications concerning $Z_b$ at present since  neither of these potentials is known  from the lattice simulations in detail.  However, it is conceivable that a mutual effect of a significantly attractive and a very mildly attractive potential could lead to a bound state closely below $B\bar B^*$ threshold, which could be related to experimental $Z_b$. Further lattice and analytical studies listed in Sec. \ref{sec:outlook} are required to reach conclusions along these lines.   
\end{itemize}
Let us note that the $Z_b(10610)$ was found as a virtual bound state slightly below the threshold by the reanalysis of the experimental data \cite{Wang:2018jlv}  (\footnote{Virtual bound state is obtained in \cite{Wang:2018jlv} when the coupling to bottomonium light-meson channels was turned off. The position of the pole is only slightly shifted when this small coupling is taken into account.}). In \cite{Wang:2018jlv,Bondar:2011ev} $Z_b(10610)$ and $Z_b(10650)$ are dominated by $B\bar B^*$ and $B^*\bar B^*$, respectively, which  suggests that potentials for components $J^l\!=\!0$ and $ J^l\!=\!1$ are of similar size\footnote{One can see this in the EFT potential for the $Z_b$ in Eq. (2.4) of Ref. \cite{Baru:2017gwo}, which is written in the $B\bar B^*$ and $B^*\bar B^*$ basis. 
The $C_1$ and $C'_1$ correspond to the potentials of the $J^l\!=\!0$ and $ J^l\!=\!1$ components. }. This conclusion is somewhat different to the conclusion of our simulation which suggests that $ J^l\!=\!0$ component is more attractive than $ J^l\!=\!1$, as shown in Fig. \ref{fig:5}. Let us point out that the lattice potentials were extracted assuming that the eigenstate dominated by $B\bar B^*$ couples only to  this Fock component and does not contain other Fock components. It remains to be explored in the future if both potentials would be of a more similar size once this assumption is relaxed.

\section{Comparison with previous lattice studies of $\bar bb \bar qq$ with $I\!=\!1$}\label{sec:comparison}
  
The list of relevant quantum numbers is given in Table \ref{tab:1}. The system with  $ C\!\cdot\! P\!=\! \epsilon\!=\!-1$ was already studied on the lattice \cite{Peters:2016wjm,Prelovsek:2019ywc} and a sizable attraction between $B$ and $\bar B^{*}$ was found, as detailed in Sec. \ref{sec:eigen-energies}. The other three quantum numbers have not been considered before, except for the ground state for  quantum channel $C\!\cdot\! P\!=\! \epsilon\!=\!+1$ that was extracted in \cite{Alberti:2016dru}. This study explored the hadroquarkonium picture, where the quarkonium could be bound inside the core of a light hadron. They calculated the difference $\Delta V_l(r)$ between the potential of a heavy quark-antiquark pair in the background of a light hadron $l$ and the potential in the vacuum. Their lattice setup enabled a better accuracy  $O({\mathrm MeV})$ on the energy for the ground state. Slightly attractive potential with the size up to few MeV was found for the majority of light hadrons, while their result is compatible with zero for $l=\rho$. This is consistent with our result, where no observable interaction is found between $\bar bb$ and $\rho$.

\section{Outlook}\label{sec:outlook}

The presented simulations of $\bar bb\bar qq$ system with $I\!=\!1$ represent only the first step towards exploring  the energy region near $m_{Zb}\simeq m_{B}+m_{B^*}$, where a number of severe simplifications have been made. It would be valuable if the future lattice simulation could determine the eigenenergies of the considered channels with  an improved accuracy. The simulations with smaller lattice spacing would be needed to extract static potentials at smaller separations between static quarks. The simulations with larger volumes would be more challenging since the discrete spectrum of $[\bar bb]l(p)$ states would be denser and more interpolators would be needed to explore the same energy region. One could extend the operator basis also with additional operator types, for example those with diquark-antidiquark structure. A much more difficult challenge would be to take into account the resonance nature of  $\rho,~b_1,~a_1$, and $a_0$ decaying to multiple hadrons, which will require implementation of multihadron operators $O_{[\bar bb] l_1(p_1) l_2(p_2) ... }$.  We note that the relation between the eigenenergies of three hadrons and their scattering amplitude has been analytically derived (for example, in \cite{Hansen:2019nir,Hammer:2017kms,Mai:2017bge}) which would be helpful in these considerations.   

On the analytical side, the derivation of  the form  for the static potential between $B^{(*)}$ and $\bar B^{(*)}$ at very small $r$ would be valuable. It is still open as to how the conclusions could be affected due to the small overlap between the eigenstate dominated by $B\bar B^*$ and $[\bar bb] [\bar qq]$ interpolators. The $Z_b$ is a linear combination of two quantum channels (\ref{decomposition}) and the current knowledge on their static potentials is shown in Fig. \ref{fig:5}. An analytic study that considers the dynamics of the  $B\bar B^*$ and $B^*\bar B^*$ channels  which are coupled via those static potentials would be needed, particularly after those potentials will be extracted with better accuracy in the future.

\section{Conclusions}\label{sec:conclusions}

Two $Z_b$ resonances with $J^P=1^+$  were the first discovered bottomoniumlike tetraquarks. They predominantly decay to $B\bar B^*$ and $B^*\bar B^*$ and lie slightly above these two thresholds. They decay also to a bottomonium and a pion, which implies the exotic quark content  $\bar bb\bar du$.  Our aim  is to explore whether the interaction between $B^{(*)}$ and $\bar B^*$ is responsible for the existence of $Z_b$. The main challenge is that  $Z_b$ decays to $\bar b u+\bar d b$ as well as lower-lying states $\bar bb+\bar du$

We study the system $\bar bb\bar du$ (\footnote{Our results for $I\!=\!1$  apply to isospin components $I_3=-1,0,1$.}) with the static $\bar{b}b$ pair separated by $r$ on the lattice. Several quantum channels are considered and operators of type $[\bar bu][\bar db]$ and $[\bar bb][\bar du]$ are employed. We determine eigenenergies $E_n(r)$ and compare them to the noninteracting energies  of two-hadron systems $\bar b u+\bar d b$ and $\bar bb+\bar du$.  If the $E_n(r)$ is below (above) noninteracting, then the state feels the attraction (repulsion).  

The  $Z_b$ with finite $m_b$ can decay to $\Upsilon \pi$ and $\eta_b \rho$ (among others), while these two quantum channels are decoupled for the static $b$ quarks used in the simulation. The simulation \cite{Prelovsek:2019ywc} considered the quantum number that couples to $\Upsilon \pi$ and found that the static potential between $B$ and $\bar B^*$ is significantly attractive at $r<\SI{0.4}{fm}$. The present simulation considers the quantum number that couples to $\eta_b \rho$ and finds that potential between $B$ and $\bar B^*$ is consistent with zero, except for a slight attraction at $r\simeq 0.1~$fm. The  first attractive potential alone leads to a bound state below $m_B+m_{B^*}$ that could be related to $Z_b$ \cite{Prelovsek:2019ywc}, but it is likely somewhat too deep. A future analytic study will be needed to determine  the mass of $Z_b$ that arises from the mutual effect of both potentials. It is conceivable that the mutual effect of both potentials could lead to a $Z_b$ state in the vicinity of the $m_B+m_{B^*}$ threshold. Section \ref{sec:outlook} lists the improvements from the lattice and analytic studies that would be valuable to obtain more solid conclusions concerning $Z_b$. 

We explore also two quantum channels  $\bar bb\bar du$ that couple only to $\bar bb+\bar du$ and we find negligible interaction between bottomonium and the light hadrons. The interaction between bottomonium and light hadrons is also found to be small for two quantum channels that do couple to $Z_b$.

\section*{Acknowledgments}

We thank  V. Baru, P. Bicudo, N. Brambilla,  T. Cohen, C. Hanhart, M. Karliner, R. Mizuk, J. Soto, A. Peters, J. Tarrus and M. Wagner for valuable discussions.
S.P. acknowledges support by Slovenian Research Agency ARRS (Research Core Funding No. P1-0035 and No. J1-8137) and DFG Grant No. SFB/TRR 55. The work of M. S. is supported by Slovenian Research Agency ARRS (Grant No. 53647).

\appendix
\section{Effective masses and overlaps}\label{sec:appendixA}

\begin{figure*}[ht!]
\centering
\hspace*{-2.3cm} 
\begin{subfigure}{0.425\textwidth}
\includegraphics[width=1.45\textwidth]{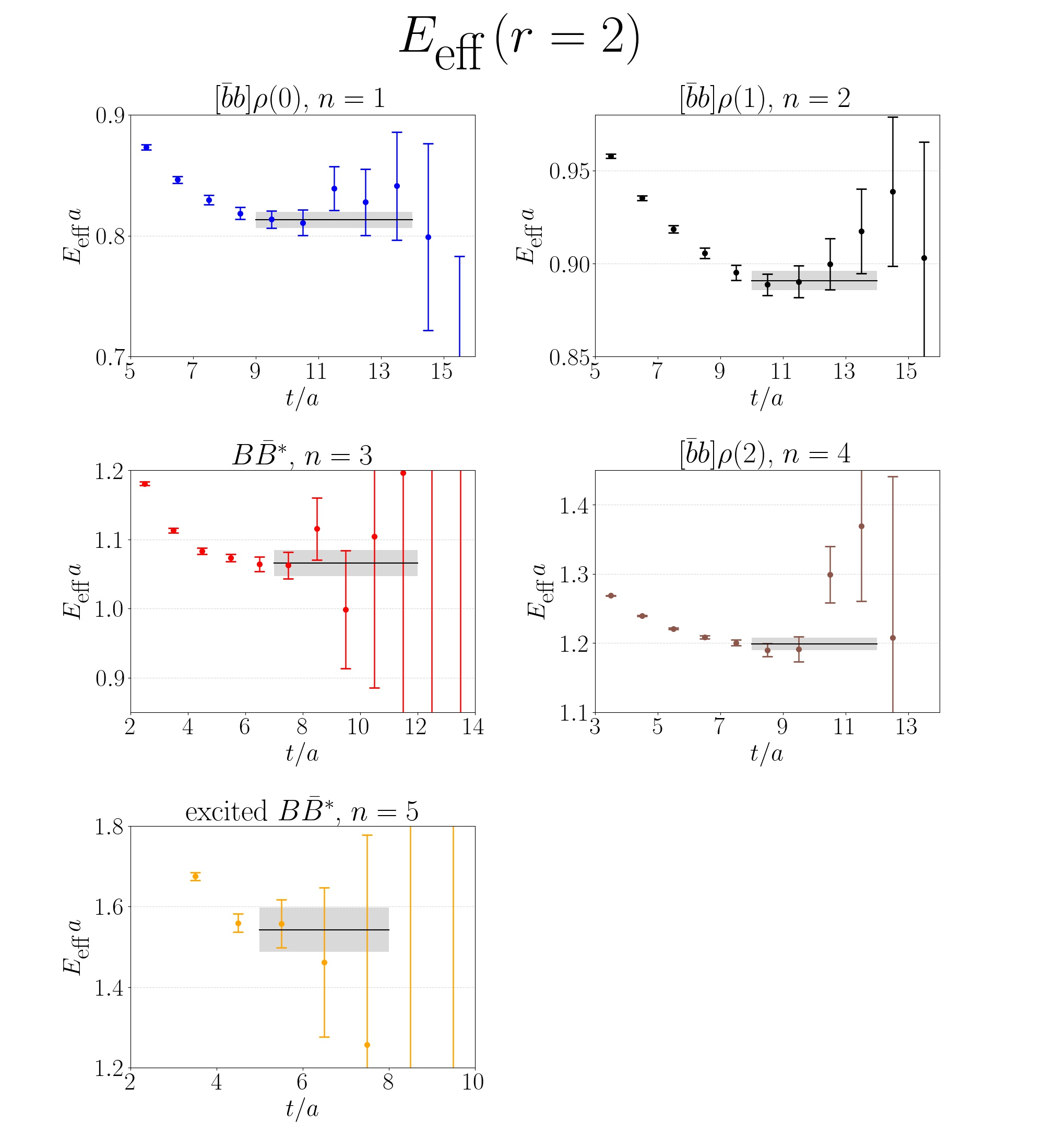} 
\caption{$C\!\cdot\! P\!=\! \epsilon\!=\!+1$} \label{fig:Eeff1a}
\end{subfigure}  
\hspace*{3cm}  
\begin{subfigure}{0.33\textwidth}
\vspace*{0.15cm}
\includegraphics[width=1.27\textwidth]{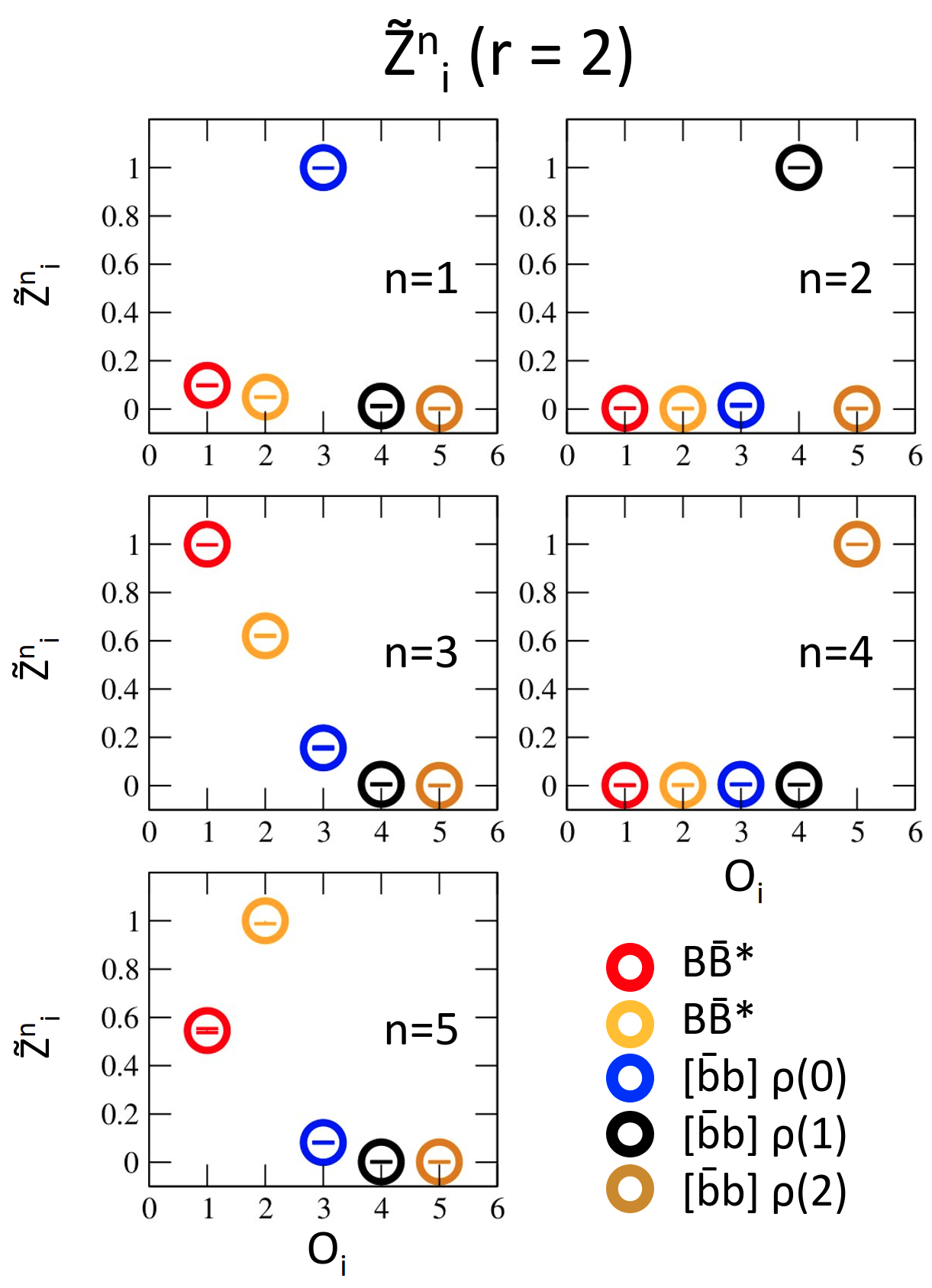}
\vspace*{1.2cm}
\caption{$C\!\cdot\! P\!=\! \epsilon\!=\!+1$} \label{fig:Eeff1b}
\end{subfigure}
\caption{  (a)  Effective energies $E_n^{\textrm{eff}}$ of the system in Fig. \ref{fig:1b} for separation $r/a=2$ and all eigenstates $n=1,\ldots,5$. They render eigenenergies $E_n$ in the plateau region. (b) Normalized overlaps $\tilde Z_i^n\propto \langle O_i|n\rangle$ of each eigenstate $n$ in (a) to the five operators \eqref{E3}. Shown are absolute values of the overlaps for $r/a=2$.}
\label{fig:Eeff1}
\end{figure*}

\begin{figure*}[h!]
\centering
\hspace*{-1.6cm} 
\begin{subfigure}{0.45\textwidth}
\includegraphics[width=1.2\textwidth]{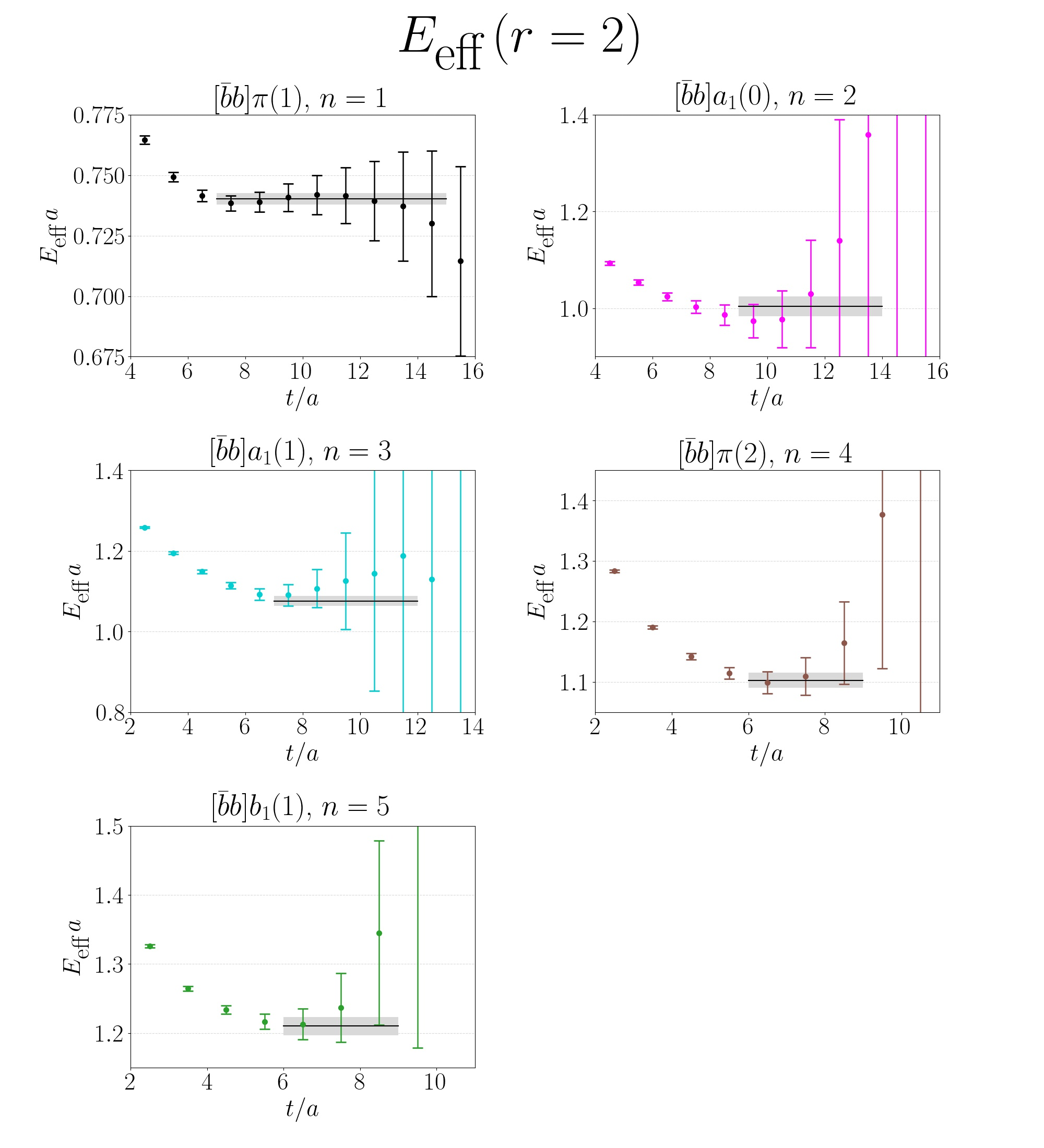}
\caption{$C\!\cdot\! P=+1,$ $\epsilon=-1$} \label{fig:Eeff2a}
\end{subfigure} 
\hspace*{0.5cm}
\begin{subfigure}{0.45\textwidth}
\vspace*{-0.15cm}
\includegraphics[width=1.2\textwidth]{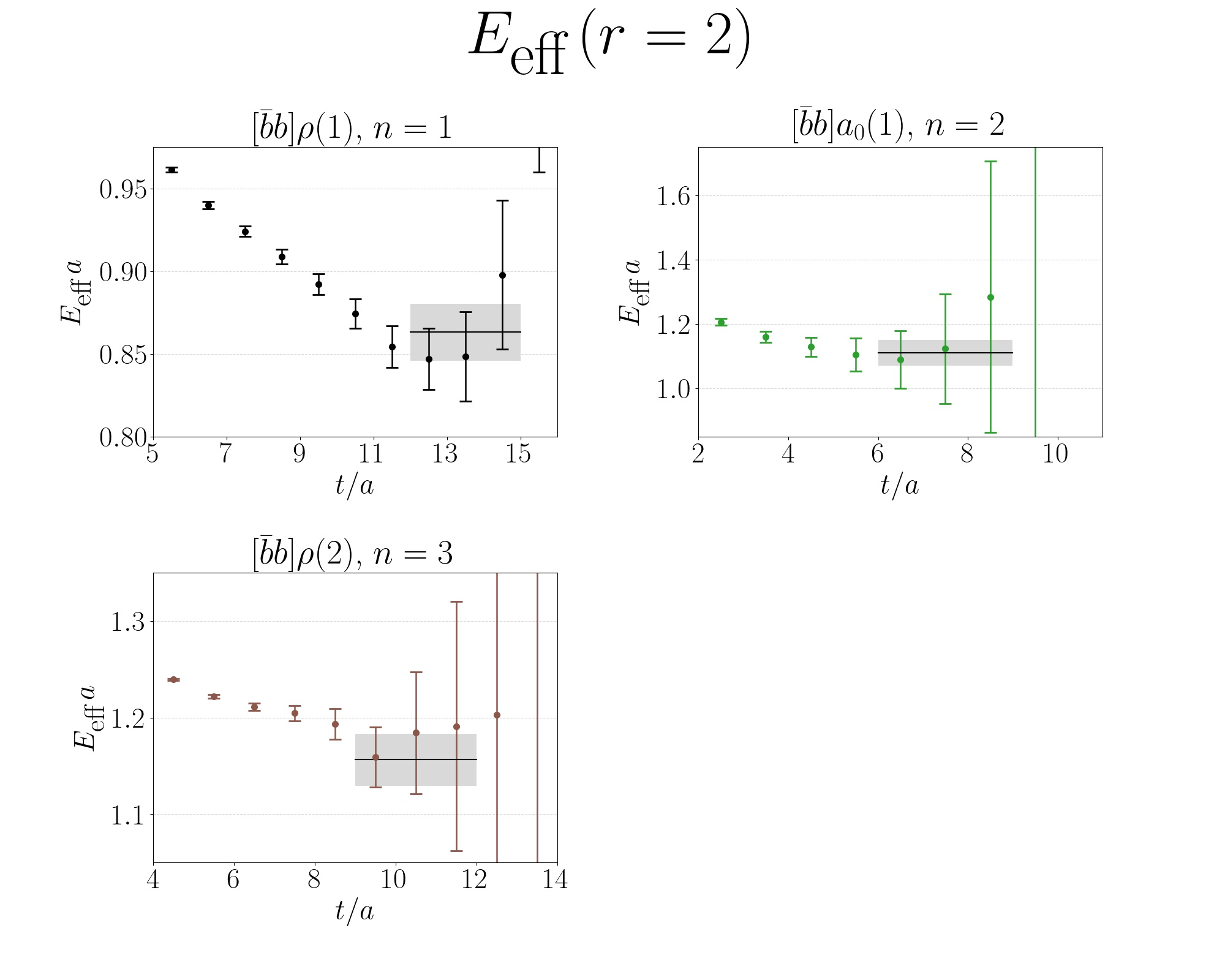}
\vspace*{2.6cm}
\caption{$C\!\cdot\! P=-1,$ $\epsilon=+1$} \label{fig:Eeff2b}
\end{subfigure}
\vspace*{-0.2cm}
\caption{Effective energies $E_n^{\textrm{eff}}$ and the fitted plateau $E_n$ at $r/a=2$.}
\label{fig:Eeff2}
\end{figure*}

\begin{figure*}[h!]
\centering
\hspace*{-1.9cm} 
\begin{subfigure}{0.42\textwidth}
\includegraphics[width=1.48\textwidth]{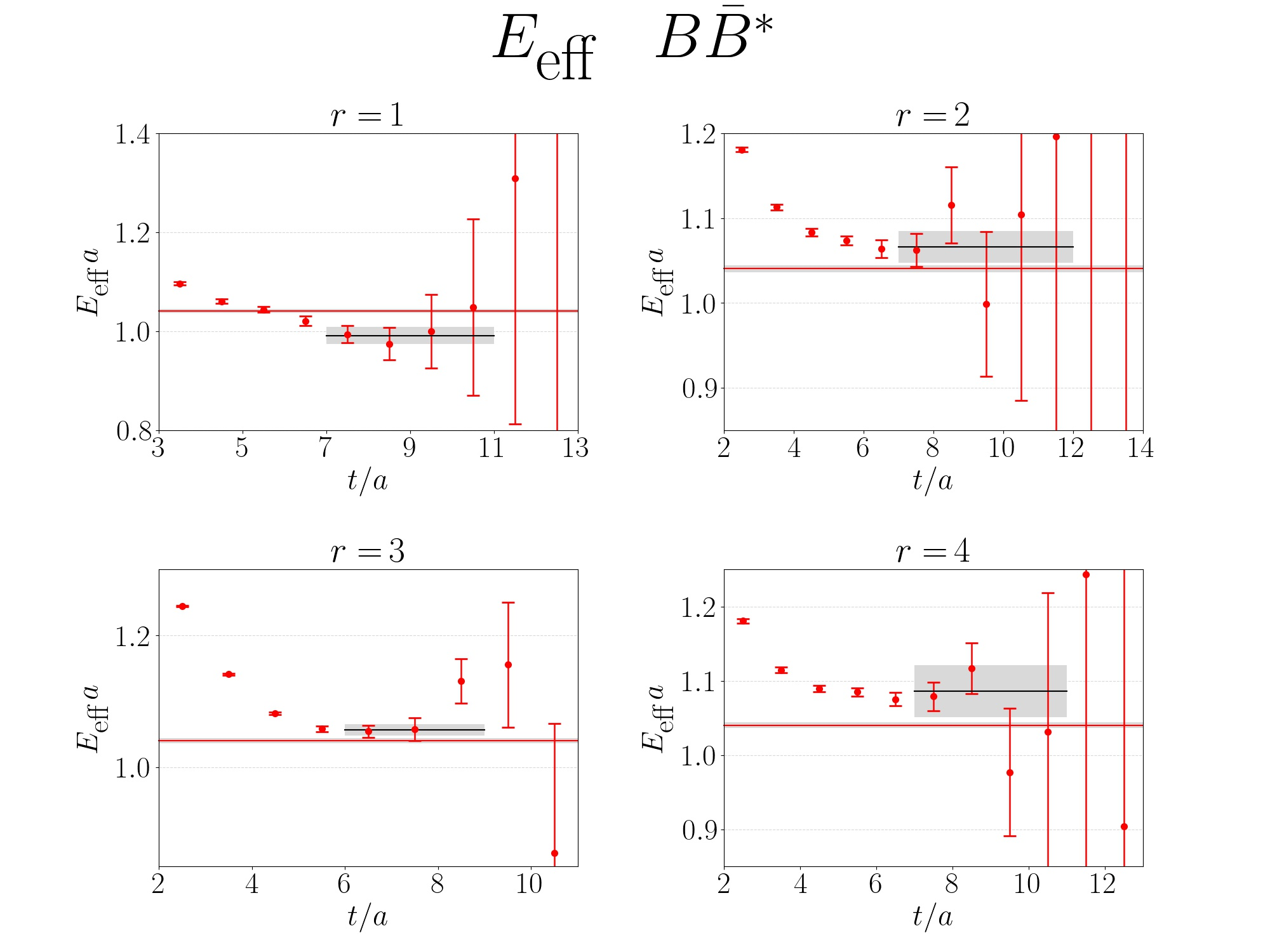}
\vspace*{0.01cm}
\caption{$C\!\cdot\! P\!=\! \epsilon\!=\!+1$} \label{fig:Eeff3a}
\end{subfigure} 
\hspace*{3.2cm}
\begin{subfigure}{0.325\textwidth}
\includegraphics[width=1.15\textwidth]{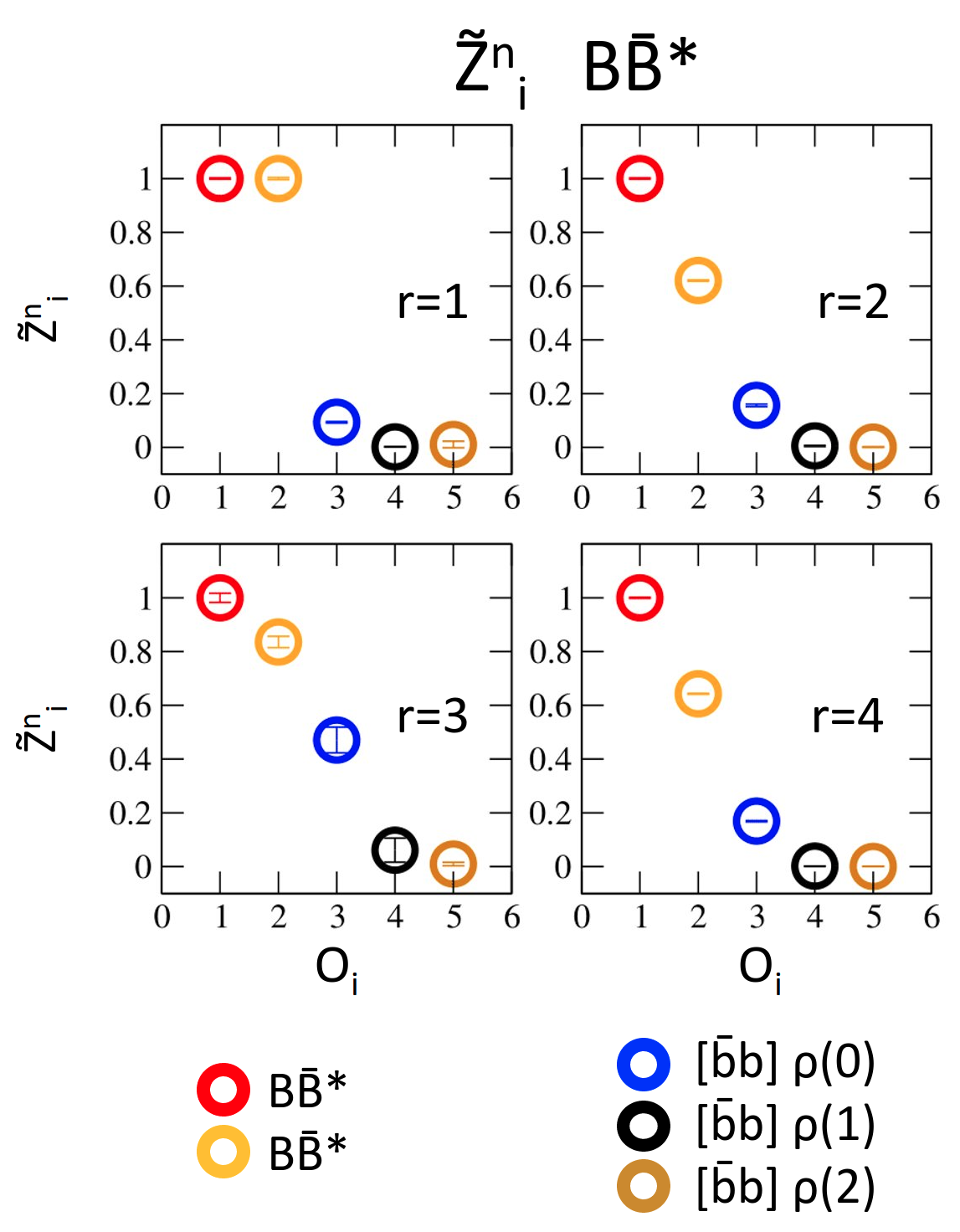}
\caption{$C\!\cdot\! P\!=\! \epsilon\!=\!+1$} \label{fig:Eeff3b}
\end{subfigure}
\caption{Eigenstate dominated by $B\bar B^*$ (red circles in Fig. \ref{fig:2}) in the quantum channel $C\!\cdot\! P\!=\! \epsilon\!=\!+1$ for separations $r/a=1,\ldots,4$: (a) effective energies $E^{\textrm{eff}}(t)$ and (b) normalized overlaps $\tilde Z_i^n\propto \langle O_i|n\rangle$, where the absolute value of the overlap is shown.}
\label{fig:Eeff3}
\end{figure*}

The effective energies $E_n^{\textrm{eff}}$ of the system in Fig. \ref{fig:1a} are shown in Figs. \ref{fig:Eeff1a}, \ref{fig:Eeff2a}, \ref{fig:Eeff2b} for the separation $r/a=2$ and all eigenstates $n=1,\ldots,5(3)$ for all three considered quantum channels. They are obtained from the correlation matrices $C_{ij}(t)$ with the variational approach $C(t)u_n(t)=\lambda_n(t)  C(t_0)u_n(t)$, where the effective energies are given by the eigenvalues $E_n^{\textrm{eff}}(t)\equiv\ln[\lambda_n(t)/\lambda_n(t+1)]$. Results for $t_0/a=2$ are shown. The effective energies render eigenenergies $E_n$ in the plateau region, indicated in the plots.

The overlaps $\langle O_i|n\rangle$ of each eigenstate $n$ to employed operators $O_i$ (see Eq. \eqref{E3}) of the quantum channel $C\!\cdot\! P\!=\! \epsilon\!=\!+1$  are shown in terms of the normalized overlaps $\tilde Z_i^n$ in Fig. \ref{fig:Eeff1b}. Here $\tilde Z_i^n\equiv \langle O_i|n\rangle/\max_m \langle O_i|m\rangle$ is normalized so that its maximal value for given $O_i$ across all eigenstates is equal to one.  

The effective energies, their fits and normalized overlaps of the eigenstate dominated by $B\bar B^*$ are presented in Fig. \ref{fig:Eeff3}.

\section{Ratio method}\label{sec:ratio}

\begin{figure}[h!]
\begin{center}
\includegraphics[width=0.53\textwidth]{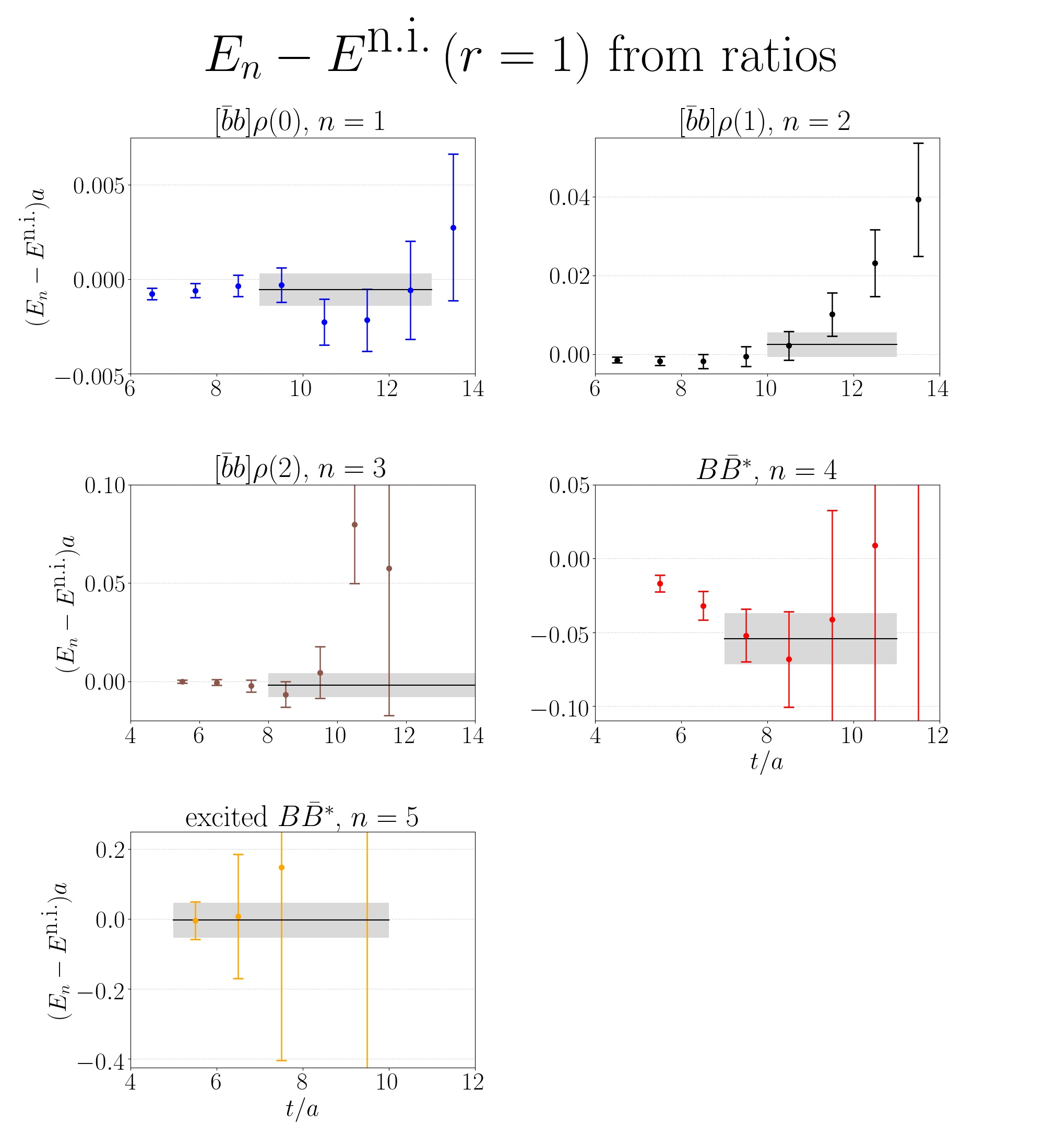}   
\caption{Effective energies for the ratio method. Shown is the quantum channel $C\!\cdot\! P\!=\! \epsilon\!=\!+1$ for $r/a=1$ and with the same color coding as in Fig. \ref{fig:Eeff1} . The plateau corresponds to the energy shift.}
\label{fig:ratio}
\end{center}
\end{figure}

Our main result is the comparison of eigenenergies and the noninteracting energies, where both are extracted using the GEVP approach. Additionally, we made a cross-check by considering the ratio
\begin{equation}
\label{ratio1} 
\frac{ \lambda_{4q}^{(n)}(t) }{ \lambda_{2q,\textrm{a}}(t)  \lambda_{2q,\textrm{b}}(t) }\> \propto  e^{ -(E_n-E^{\textrm{n.i.}})t}
\end{equation}
where $\lambda_{4q}^{(n)}(t)$ is the $n$th GEVP eigenvalue of the correlation matrices constructed with \eqref{E3}, \eqref{E4}, \eqref{E5}, and $ \lambda_{2q,\textrm{a,b}}(t)$ are the eigenvalues of the noninteracting part, where we chose operators with the largest overlap to the $n$th state. The effective energies for the ratio with fits are shown in Fig. \ref{fig:ratio}. 

 \end{document}